\documentclass[a4paper,fleqn]{cas-sc}

\usepackage[authoryear]{natbib}

\usepackage{lineno}
\usepackage{amssymb}
\usepackage{amsmath}
\usepackage{color}
\usepackage{soul}
\usepackage{ulem}
\usepackage{float}
\usepackage{caption}
\begin{document}
\let\WriteBookmarks\relax
\def\floatpagepagefraction{1}
\def\textpagefraction{.001}
\shorttitle{}
\shortauthors{Y. Ye et~al.}
\let\printorcid\relax

\title [mode = title]{Nonreciprocal scattering of elastic waves at time interfaces induced by spatiotemporal modulation}                      



\author[a,b]{Yingrui Ye}[]
\author[c]{Chunxia Liu}[]
\author[b]{Alessandro Marzani}[]
\author[d]{Emanuele Riva}[]
\author[b]{Antonio Palermo}[]\cormark[1]\ead{antonio.palermo6@unibo.it}
\author[a]{Xiaopeng Wang}[]\cormark[1]\ead{xpwang@mail.xjtu.edu.cn}

\cortext[cor1]{Corresponding author}

\affiliation[a]{organization={Department of Mechanical Engineering, Xi'an Jiaotong University},
            city={Xi'an},
            postcode={710049}, 
            country={China}}
\affiliation[b]{organization={Department of Civil, Chemical, Environmental and Materials Engineering, University of Bologna},
            city={Bologna},
            postcode={40136}, 
            country={Italy}}            
\affiliation[c]{organization={College of Electrical Engineering and Automation, Anhui University},
            city={Hefei},
            postcode={230601}, 
            country={China}}
\affiliation[d]{organization={Department of Mechanical Engineering, Politecnico di Milano},
            city={Milano},
            postcode={20156}, 
            country={Italy}}

\begin{abstract}
Spatiotemporally modulated elastic metamaterials have garnered increasing interest for their potential applications in nonreciprocal wave devices. Most existing studies, however, focus on systems where spatiotemporal modulation is continuous and infinite in time. Here, we investigate the temporal dynamics of elastic waves at time interfaces created by the sudden activation or deactivation of spatiotemporal modulation in a medium's elastic properties. By developing an ad hoc mode-coupling theory, we reveal that such time interfaces enable controlled frequency and wavenumber conversion through mode redistribution and energy pumping. Specifically, we quantitatively evaluate the temporal scattering behavior of elastic longitudinal waves under two representative spatiotemporal modulations: subsonic and supersonic. These modulations give rise to frequency and wavenumber bandgaps, respectively. We demonstrate that subsonic modulation induces nonreciprocal energy reversal, while supersonic modulation leads to nonreciprocal energy amplification. Our findings pave the way for the development of temporal elastic metamaterials with practical applications in designing one-way elastic filters, amplifiers, and frequency converters.
\end{abstract}


\begin{highlights}
\item We propose a mode-coupling theory to investigate the temporal scattering mechanism. 
\item Bandgaps evolution of an elastic spatiotemporally modulated medium is discussed.
\item Nonreciprocal energy reversal for subsonic modulation is demonstrated.
\item Stable nonreciprocal parametric amplification for supersonic modulation is demonstrated.
\end{highlights}


\maketitle


\section{Introduction}
\label{sec1}
The study of wave propagation phenomena in time-varying media has recently seen a renewed interest in both the physics and engineering communities \citep{miyamaru2021ultrafast,galiffi2022photonics,engheta2023four,vazquez2023incandescent,horsley2023quantum,wapenaar2025green}. This interest is primarily driven by the potential to unveil uncommon dynamic features, such as topological insulation \citep{li2019topological,wang2021topological,ni2021topological}, temporal pumping \citep{Xia2021experimental}, temporal aiming \citep{santini2023elastic}, temporal disorder \citep{carminati2021universal}, frequency shift \citep{wu2019serrodyne,ramaccia2019phase}, and wave amplification \citep{koutserimpas2018nonreciprocal,lyubarov2022amplified,dikopoltsev2022light}. When time modulation varies along the waveguide, producing a wave-like spatiotemporal modulation of the constitutive properties, nonreciprocity may arise from the breaking of time-reversal symmetry \citep{sounas2017non,nassar2017modulated,nassar2020nonreciprocity}. 
Wave propagation in spatiotemporally modulated media was initially investigated in circuit systems \citep{cassedy1963dispersion,cassedy1967dispersion} and later systematically extended to optics \citep{chamanara2017optical,pendry2021gain} and acoustics \citep{fleury2015subwavelength,wang2018observation,li2019nonreciprocal,chen2024experimental}. In elastic systems, this has recently been realized by dynamically varying the elastic or inertial properties of the host medium \citep{trainiti2016non,goldsberry2020nonreciprocal,Marconi2020experimental} or by utilizing modulated resonators \citep{nassar2017non,chen2019nonreciprocal,wu2021non,palermo2020surface,CELLI2024}. Through spatiotemporal modulation, a range of exotic elastic wave propagation phenomena have been demonstrated, including nonreciprocity \citep{trainiti2016non,goldsberry2020nonreciprocal,nassar2017non,wu2021non,palermo2020surface,Marconi2020experimental,chen2019nonreciprocal} and frequency conversion \citep{nassar2017non,wu2021non,palermo2020surface,Marconi2020experimental,chen2019nonreciprocal}.

Still, most investigations focus on temporally continuous and infinite modulations, overlooking the temporal scattering phenomena associated with the activation and deactivation of the spatiotemporal modulation in time. Indeed, a comprehensive understanding of the dynamics of such modulated systems requires the development of wave-scattering theory capable of describing the dynamics at time interfaces. At a time interface, the temporal properties of waves (i.e. frequency and group velocity) change abruptly, while the spatial property (i.e. wavenumber) remains unchanged \citep{xiao2014reflection,lee2018linear,zhou2020broadband,wapenaar2024waves}; this is in contrast to the frequency-preserving energy transfer associated with space interfaces.
These aspects, well analysed in the context of electromagnetic waves \citep{Moussa2023observation,jones2024time},  have only been recently discussed and experimentally observed in time-varying discrete mechanical systems, namely phononic crystals of discrete repelling magnets \citep{kim2024temporal}, and for transverse waves propagating in an elastic strip with a moving interface \citep{delory2024elastic}. More recently, a detailed account of temporal reflection and refraction was provided for flexural waves propagating in modulated mechanical metabeams, although restricted to media with only time-varying properties \citep{wang2025temporal}.

In this study, we propose a theoretical framework to model the temporal scattering of elastic waves across time interfaces in a spatiotemporally modulated medium. Our focus is on the dynamic behavior of elastic longitudinal waves propagating through a spatiotemporally modulated temporal interlayer bounded by two time interfaces. We develop a mode-coupling theory that captures the mechanisms of frequency conversion and energy transport induced by temporal interlayers under subsonic and supersonic modulation regimes. Notably, these two regimes exhibit distinct dispersive characteristics, with either frequency or wavenumber bandgaps populating the dispersion curves. Numerical simulations validate our theoretical findings and demonstrate a range of unique nonreciprocal wave phenomena, including one-way filtering, parametric amplification, and frequency conversion.

The paper is organized as follows. In \textcolor{blue}{Section \ref{sec2}}, we describe the dispersion relations of an elastic modulated medium with different modulation velocities and derive the mode-coupling theory based on the mode redistribution and degeneracy at time interfaces. In \textcolor{blue}{Section \ref{sec3}}, we employ our theoretical framework to calculate the temporal scattering behavior for a subsonic modulation case, and then verify the nonreciprocal scattering using numerical simulations. In \textcolor{blue}{Section \ref{sec4}}, we further analyze the nonreciprocal temporal scattering for a supersonic modulation case analytically and numerically. Finally, we present conclusions and prospects in \textcolor{blue}{Section \ref{sec5}}.

\section{Theoretical model}
\label{sec2}
In this section, we establish a theoretical framework to describe the scattering due to temporal interfaces in an elastic waveguide with spatiotemporal modulation. To this purpose, in what follows, we begin by briefly recalling the dispersion relations of longitudinal waves under spatiotemporal stiffness modulation.

\subsection{Longitudinal wave propagation across time interfaces}
\label{subsec2.1}
\begin{figure}[h]   \centering
\includegraphics[width=1\linewidth]{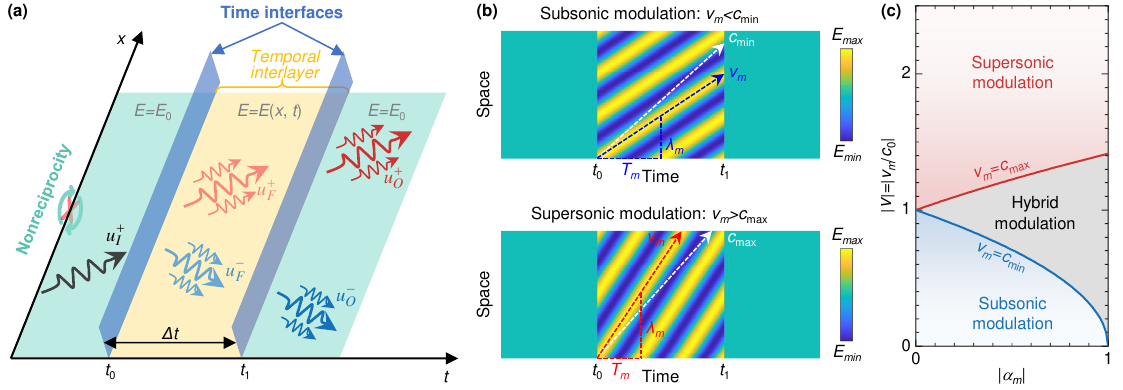}
\caption{Schematic of a temporal elastic medium with spatiotemporally modulated stiffness $E(x,t)$. (a) Schematic of temporal scattering of elastic longitudinal waves crossing a modulated temporal interlayer ($t_{0} \leq t \leq t_{1}$). The modulation duration is $\Delta t=t_1-t_0$. (b) Contours of elastic modulus for subsonic and supersonic modulation. In the temporal interlayer, $E(x,t)$ is space-time modulated in a cosine pump wave. The propagation of guided waves and subsonic (supersonic) pump waves are depicted by white and blue (red) dashed arrows, and their slopes represent the corresponding phase velocities. (c) Modulation regimes for different dimensionless modulation velocities $V$ and amplitudes $|\alpha_{m}|$.}
\label{Fig.1}
\end{figure}

We consider the propagation of 1D longitudinal waves along an elastic rod subjected to a spatiotemporal modulation occurring in the time window $[t_0, t_1]$. The instants $t_0$ and $t_1$ mark the occurrence of time interfaces, while the overall time window defines the spatiotemporally modulated interlayer (see \textcolor{blue}{Fig. \ref{Fig.1}}(a)). Longitudinal waves in a rod or slender beam, where rotational inertia and shear deformation are negligible, are governed by:
\begin{equation}\label{Eq.1}
\partial_{x}\sigma-\partial_{t}(\rho v)=0,
\end{equation}
where, $\sigma=E\partial_{x}u$ is the normal stress, $v=\partial_{t}u$ is the velocity of the particle, $E$ and $\rho$ denote the Young’s modulus and mass density. The longitudinal displacement along the $x$-direction is described by $u=u(x,t)$. The operator $\partial_{i}=\frac{\partial}{\partial i}$ with $i$ denoting the space ($x$) or time ($t$) coordinates is a partial derivative operator with respect to $i$.

When space-time-dependent constitutive properties are considered, Eq. (\textcolor{blue}{\ref{Eq.1}}) is reformulated as:
\begin{equation}\label{Eq.2}
\partial_{x}\Big[E(x,t)\partial_{x}u(x,t)\Big]-\partial_{t}\Big[\rho(x,t)\partial_{t}u(x,t)\Big]=0.
\end{equation}

As anticipated, the modulation is active within the time window $[t_0, t_1]$. As such, the constitutive properties remain constant for $t<t_{0}$ and $t>t_{1}$, while varying with both time and space within temporal interlayer ($t_{0} \leq t \leq t_{1}$), as shown in \textcolor{blue}{Fig. \ref{Fig.1}}(a). Notably, modulated elastic properties are easily implemented in mechanical waveguides, leveraging, for example, electromechanical coupling based on piezoelectric elements \citep{Trainiti2019time,Marconi2020experimental,Xia2021experimental}. Hence, in what follows, we assume a modulated elastic modulus, while the mass density is kept uniform and constant, i.e., $\rho(x,t)=\rho_{0}$. For the elastic modulus, we consider the following modulation laws:
\begin{equation}\label{Eq.3}
E(x,t)=
\begin{cases}
E_{0}, & t<t_{0}, \\
E_{0}[1+\alpha_{m}\cos(\omega_{m}t-\kappa_{m}x)], & t_{0} \leq t \leq t_{1}, \\
E_{0}, & t>t_{1},
\end{cases}
\end{equation}
where $E_{0}$ is the non-modulated elastic modulus, $\alpha_{m}$ is the normalized modulation amplitude, and $\omega_{m} = 2\pi/T_{m}$ and $\kappa_{m} = 2\pi/\lambda_{m}$ are the modulation angular frequency and wavenumber, with $T_{m}$ and $\lambda_{m}$ being the corresponding temporal and spatial periods, respectively.

Notably, the wave velocity in the modulated interlayer is not a constant value, but varies within a range:
\begin{equation}\label{Eq.4}
c_{\min}<c=\sqrt{\frac{E(x,t)}{\rho_0}}<c_{\max},
\end{equation}
where $c_{\min}=\sqrt{1-|\alpha_{m}|}c_{0}$ and $c_{\max}=\sqrt{1+|\alpha_{m}|}c_{0}$ are the minimum and maximum wave velocities in the modulated interlayer, and $c_{0}=\sqrt{E_{0}/\rho_{0}}$ is the longitudinal wave velocity in the homogeneous medium.

The modulation pattern takes the form of a traveling wave, with a dimensionless modulation velocity given as
$V={v_{m}}/{c_{0}}$, where $v_{m}={\omega_{m}}/{\kappa_{m}}$ is the modulation phase velocity. In \textcolor{blue}{Fig. \ref{Fig.1}}(b), we display the elastic modulus contours in the space-time domain for two representative modulation patterns, i.e., subsonic and supersonic modulation.

The dispersive properties of Floquet-Bloch waves traveling through the spatiotemporally modulated interlayer depend on the modulation velocity and amplitude. We consider modulation functions with different dimensionless modulation amplitudes $\alpha_m$ and dimensionless modulation velocities $V$. Accordingly, subsonic and supersonic modulation regimes can be identified by $v_m<c_{\min}$ and $v_m>c_{\max}$, as shown in \textcolor{blue}{Fig. \ref{Fig.1}}(c). These two regimes are separated by a hybrid modulation region (sonic region), where the modulation velocity $v_m$ can be either faster or slower than the wave velocity in the modulated interlayer, $c$, depending on its fluctuation induced by spatiotemporal modulation. We remark that a supersonic modulation yields an unstable interaction between the pump waves and the guided waves, ultimately resulting in an unbounded, time-growing amplitude in a lossless and continuously modulated medium \citep{cassedy1967dispersion,trainiti2016non}. However, we expect such amplification to remain bounded when a finite spatiotemporal modulation is considered, as shown in the following.

\subsection{Dispersion relation of spatiotemporally modulated media}
\label{subsec2.2}
To support our investigation on the temporal scattering of elastic waves at the time interface of a spatiotemporally modulated interlayer, we briefly recall the dispersive properties of longitudinal waves passing through the unbounded homogeneous and modulated media. In the homogeneous medium with constant constitutive properties ($\rho_0$, $E_0$), by assuming $u(x,t)=\hat{u}_{0}e^{i(\omega t-\kappa x)}$ as the displacement solution and substituting it into the motion equation shown in \textcolor{blue}{Eq. (\ref{Eq.2})}, the dispersion relation is simply obtained as:
\begin{equation}\label{Eq.6}
\omega(\kappa)=\pm c_{0}\kappa,
\end{equation}
where the $+/-$ sign indicates the wave propagating in the forward/backward direction. 

In the modulated medium with modulated elastic stiffness $E(x,t)$, the dispersion law is derived and solved using a plane wave expansion approach, assuming that a complete solution to \textcolor{blue}{Eq. (\ref{Eq.2})} can be expressed in the generalized Floquet form:
\begin{equation}\label{Eq.7}
u(x,t)=e^{i(\omega t-\kappa x)}\sum_{n=-\infty}^{+\infty}\hat{u}_{n}e^{in(\omega_m t-\kappa_m x)},
\end{equation}
where $\hat{u}_{n}$ represents the amplitude of the $n^{\mathrm{th}}$-order Floquet-Bloch modes.

Hence, the dispersion relation for spatiotemporal stiffness-modulated media can be obtained as:
\begin{equation}\label{Eq.8}
\text{det}
\begin{bmatrix}\cdots&\cdots&\cdots&\cdots&\cdots\\
\cdots&\beta_{-1,-1}-\zeta_{-1}&\beta_{-1,0}&\beta_{-1,1}&\cdots\\
\cdots&\beta_{0,-1}&\beta_{0,0}-\zeta_{0}&\beta_{0,1}&\cdots\\
\cdots&\beta_{1,-1}&\beta_{1,0}&\beta_{1,1}-\zeta_{1}&\cdots\\
\cdots&\cdots&\cdots&\cdots&\cdots\\
\end{bmatrix}=0,
\end{equation}
where $\beta_{m,n}=(\kappa+m\kappa_{m})(\kappa+n\kappa_{m})\hat{E}_{m-n}$ and $\zeta_{n}=\rho_0(\omega+n\omega_{m})^2$ with $m,n=0,\pm1,\pm2,\cdots, \pm N$, $N$ is a desired truncation order of the plane wave expansion.

As anticipated, when a harmonic wave ($\kappa$, $\omega$) passes through a time interface, its wavenumber is preserved while the frequency is modulated. Accordingly, we solve the dispersion law by providing real wavenumbers $\kappa$ in input and solving in terms of angular frequencies $\omega$. As a result, for a given $\kappa$, we obtain $4N+2$ eigenvalues $\hat{\omega}_{s}^{+/-}$ and $4N+2$ eigenvectors $\mathbf{\hat{U}^{+/-}_s}=\left[\begin{matrix}{\hat{u}^{+/-}_{s,-N},}&{\cdots,}&{\hat{u}^{+/-}_{s,N}}\end{matrix}\right]^{T}$ ($s=-N, \cdots, +N$), where the superscript $+/-$ represents the forward/backward Floquet-Bloch modes. More details of the derivation are presented in \textcolor{blue}{Appendix \ref{App.A}}.

For our computations, we select a truncation order of $N=3$. In what follows, to generalize our discussion, we introduce the dimensionless wavenumber $\mu$ and frequency $\mathit{\Omega}$, defined as follows:
\begin{equation}\label{Eq.9}
\mu=\frac{\kappa}{\kappa_{m}},\quad\mathit{\Omega}=\frac{\omega}{c_{0}\kappa_{m}}.
\end{equation}

\begin{figure}[h]   \centering
\includegraphics[width=0.73\linewidth]{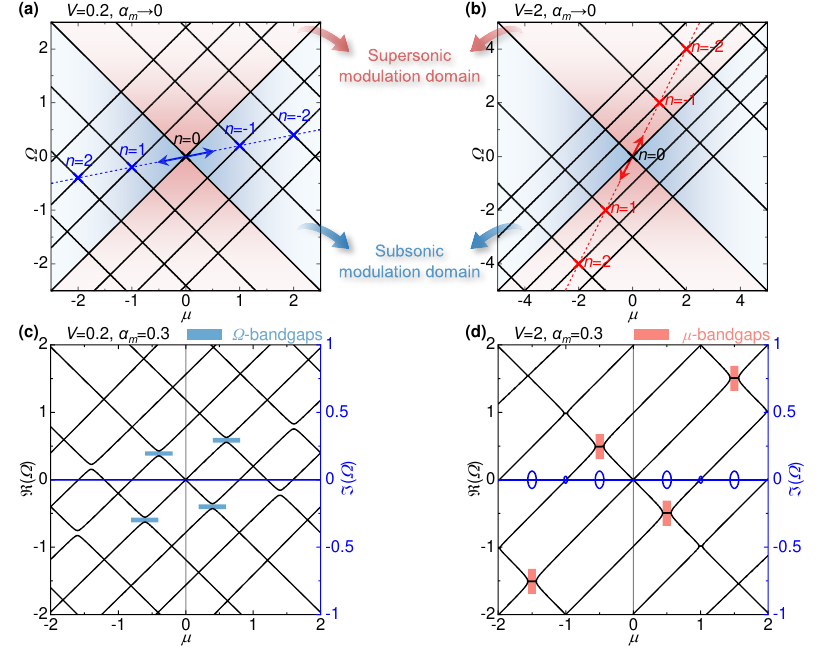}
\caption{Dispersion diagrams for longitudinal waves in spatiotemporally modulated media. (a) and (b) $\alpha_{m}\rightarrow0$, dispersion diagrams under (a) subsonic ($V=0.2$) and (b) supersonic ($V=2$) modulation, respectively. The blue and red dashed lines indicate the periodic directions of the dispersion branches, and their slopes represent the corresponding dimensionless modulation velocities $V$. The blue and red shadings correspond to the modulation domains for subsonic and supersonic modulation. (c) and (d) $\alpha_{m}=0.3$, dispersion diagrams under (c) subsonic ($V=0.2$) and (d) supersonic ($V=2$) modulation, respectively. The blue and red rectangular shadows depict the $\mathit{\Omega}$-bandgaps and $\mu$-bandgaps on the fundamental branch, respectively.}
\label{Fig.2}
\end{figure}

The dispersion diagrams for subsonic ($V=0.2$) and supersonic ($V=2$) modulation with $\alpha_{m}\rightarrow0$ and $\alpha_m = 0.3$ are shown in \textcolor{blue}{Figs. \ref{Fig.2}}(a)-(d), respectively, in a convenient four-quadrant representation adopted to better describe the temporal scattering at the time interface. The limit case for $\alpha_{m}\rightarrow0$, as shown in \textcolor{blue}{Figs. \ref{Fig.2}}(a) and (b), allows us to easily visualize how the irreducible Brillouin zone of spatiotemporal modulated media is stretched along an inclined direction. In \textcolor{blue}{Figs. \ref{Fig.2}}(c) and (d), we show the corresponding dispersion diagrams under a modulation amplitude of $\alpha_{m}=0.3$. In the presence of modulation amplitude, the dispersion curves for subsonic and supersonic modulation scenarios exhibit two distinct types of asymmetric bandgaps: frequency bandgaps ($\mathit{\Omega}$-bandgaps) and wavenumber bandgaps ($\mu$-bandgaps). For subsonic modulation, adjacent frequency bands open up to generate $\mathit{\Omega}$-bandgaps with no wave solutions. In contrast, for supersonic modulation, the adjacent frequency bands merge to generate $\mu$-bandgaps with conjugate complex solutions $\mathit{\Omega}=\Re(\mathit{\Omega})+i\Im(\mathit{\Omega})$, where the positive/negative imaginary part $\Im(\mathit{\Omega})$ represents time-decaying/time-growing waves. A detailed discussion on the transition from $\mathit{\Omega}$-bandgaps to $\mu$-bandgaps as the modulation velocity increases is provided in \textcolor{blue}{Appendix \ref{App.B}}.

According to the geometric relationship between the periodic branches, we can obtain the central wavenumber and frequency of $\mathit{\Omega}$-/$\mu$-bandgaps on the fundamental branch. In both cases, the forward directional bandgaps on the positive subbranch emerge around:
\begin{equation}\label{Eq.10}
\begin{aligned}
&\mu_{bandgap}^{+} =\pm\frac{1+V}{2},\quad \mathit{\Omega}_{bandgap}^{+}=\pm\frac{1+V}{2},
\end{aligned}
\end{equation}
while the backward directional bandgaps on the negative subbranch appear around:
\begin{equation}\label{Eq.11}
\begin{aligned}
&\mu_{bandgap}^{-} =\pm\frac{1-V}{2},\quad \mathit{\Omega}_{bandgap}^{-}=\mp\frac{1-V}{2}.
\end{aligned}
\end{equation}

\subsection{Modeling of temporal scattering via mode-coupling theory}
\label{subsec2.3}
Here we present a theoretical framework to describe the scattering mechanism of elastic longitudinal waves passing through time interfaces bounding a space-time modulated interlayer.

\begin{figure}[h]   \centering
\includegraphics[width=1\linewidth]{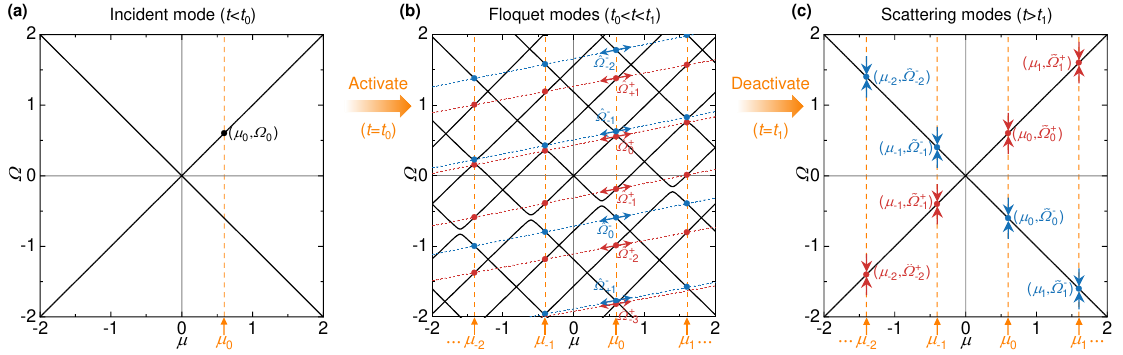}
\caption{Schematic diagram of the mode redistribution at time interfaces. (a) For $t<t_0$: the incident mode is shown as a black dot. (b) For $t_0<t<t_1$: the red and blue arrows are representative of the expansion of basic modes into positive and negative propagating Floquet modes. The positive and negative propagating Floquet modes are shown as the red and blue dots, respectively, where $\mathit{\mu_n}=\mathit{\mu_0}\pm n$. (c) For $t>t_1$: the red and blue arrows are representative of the degeneration of positive and negative propagating Floquet modes into temporal transmission and reflection modes. The positive and negative scattering modes are shown as the red and blue dots, respectively.}
\label{Fig.3}
\end{figure}

For $t<t_0$, the waveguide exhibits constant constitutive properties. In this medium, we consider a wave mode, i.e., the incident mode, described by the wavenumber-frequency pair $(\kappa_0,\omega_0)$ (or equivalently $(\mu_0,\mathit{\Omega}_0)$) and marked by a black dot on the dispersion diagram of the homogeneous medium, as shown in \textcolor{blue}{Fig. \ref{Fig.3}}(a). Thus, the incident wave takes the form of:
\begin{equation}\label{Eq.12}
u_{I}^{+}=A_{0}e^{i(\omega_{0}t-\kappa_{0}x)},
\end{equation}
where $A_0$ is the wave amplitude.

At $t=t_0$, the spatiotemporal modulation is activated, causing the incident wave to pass through the first time interface and undergo temporal scattering. As a result of the wavenumber-preserving energy transfer associated with time interfaces, a set of basic modes with the same wavenumber $\kappa_0$ ($\mu_0$) but modulated frequencies $\omega_0\to\hat{\omega}_{s}^{+/-}$($\mathit{\Omega}_0\to\hat{\mathit{\Omega}}_{s}^{+/-}$) are excited on the dispersion diagram of the spatiotemporally modulated medium, as shown in \textcolor{blue}{Fig. \ref{Fig.3}}(b). Here, we consider the scenario of subsonic modulation with $V=0.2$ and $\alpha_m = 0.3$, although the analysis is also applicable to other modulation regimes. Based on the signs of the group velocity, we label these basic modes into positive wave modes $\hat{\omega}_{s}^{+}$ ($\hat{\mathit{\Omega}}_{s}^{+}$) and negative wave modes $\hat{\omega}_{s}^{-}$ ($\hat{\mathit{\Omega}}_{s}^{-}$). For $t_0<t<t_1$, due to the pumping effect of the spatiotemporal modulation, each basic mode expands into a series of Floquet modes along a direction with a slope equal to the dimensionless modulation velocity $V$. The components of these Floquet modes are frequency-shifted by $n\omega_m$ and wavenumber-shifted by $n\kappa_m$ from the basic modes. On the dispersion diagram of the spatiotemporally modulated medium, we identify forward and backward Floquet modes by their wavenumber-frequency pairs ($\kappa_0\pm n\kappa_m,\hat{\omega}_{s}^{+/-}\pm n\omega_m$) (in dimensionless form $(\mu_0\pm n,\hat{\mathit{\Omega}}_{s}^{+/-}\pm nV)$), with $s,n=-N, \cdots, +N$,  as indicated by the red and blue dots in \textcolor{blue}{Fig. \ref{Fig.3}}(b), respectively. Consequently, the scattering effect of the time interface results in a redistribution of the incident energy into multiple groups of Floquet modes. The forward and backward Floquet waves in the modulated temporal interlayer can be expressed as:
\begin{equation}\label{Eq.13}
u_{F}^{+} =\sum_{s=-\infty}^{+\infty}C_{s}\sum_{n=-\infty}^{+\infty}\hat{u}_{s,n}^{+}e^{i[(\hat{\omega}_{s}^{+}+n\omega_{m})t-(\kappa_{0}+n\kappa_{m})x]}, 
\end{equation}
\begin{equation}\label{Eq.14}
u_{F}^{-} =\sum_{s=-\infty}^{+\infty}D_{s}\sum_{n=-\infty}^{+\infty}\hat{u}_{s,n}^{-}e^{i[(\hat{\omega}_{s}^{-}+n\omega_{m})t-(\kappa_{0}+n\kappa_{m})x]}, 
\end{equation}
where $C_s$ and $D_s$ are the amplitude coefficients of the forward and backward Floquet modes.

At $t=t_1$, the spatiotemporal modulation is deactivated, causing all Floquet modes to degenerate into temporal transmitted and reflected modes with the preserved wavenumbers $\kappa_0 \pm n\kappa_m$ ($\mu_0 \pm n$) and modulated frequencies $\hat{\omega}_{s}^{+/-} \pm n\omega_m \to \tilde{\omega}_{n}^{+/-}$ ($\hat{\mathit{\Omega}}_{s}^{+/-} \pm nV \to \tilde{\mathit{\Omega}}_{n}^{+/-}$). For $t>t_1$, we denote these temporal transmitted and reflected modes by their wavenumber-frequency pairs $(\kappa_0 \pm n\kappa_m, \tilde{\omega}_{n}^{+/-})$ (in dimensionless form $(\mu_0 \pm n, \tilde{\mathit{\Omega}}_{n}^{+/-})$), with $n=-N, \cdots, +N$, on the dispersion diagram of the homogeneous medium, as depicted by the red and blue dots in \textcolor{blue}{Fig. \ref{Fig.3}}(c).
Therefore, the scattering effect of the second time interface manifests as the degeneration of Floquet modes to temporal scattering modes. Consequently, the temporal transmission and reflection waves can be expressed as:
\begin{equation}\label{Eq.15}
u_{O}^{+}=\sum_{n=-\infty}^{+\infty}\tilde{T}_{n}e^{i[\tilde{\omega}_{n}^{+}t-(\kappa_{0}+n\kappa_{m})x]},
\end{equation}
\begin{equation}\label{Eq.16}
u_{O}^{-}=\sum_{n=-\infty}^{+\infty}\tilde{R}_{n}e^{i[\tilde{\omega}_{n}^{-}t-(\kappa_{0}+n\kappa_{m})x]},
\end{equation}
where $\tilde{T}_n$ and $\tilde{R}_n$ are the amplitudes of the $n^\mathrm{th}$-order temporal transmission and reflection waves, respectively. $\tilde{\omega}_{n}^{+}$ and $\tilde{\omega}_{n}^{-}$ denote the angular frequencies of the $n^\mathrm{th}$-order temporal transmission and reflection waves, respectively, and are given by:
\begin{equation}\label{Eq.17}
\tilde{\omega}_{n}^{+/-}=\pm c_{0}(\kappa_{0}+n\kappa_{m})
\end{equation}

To formally set up and solve the temporal scattering formalism described above, we impose displacement and momentum continuity conditions at the time interfaces. For simplicity, we set $t_0$ as the origin of time coordinates, meaning $t_0=0$ and $t_1=\Delta t$.

The continuity conditions for the time interface at $t=t_0$ can be expressed as follows:
\begin{equation}\label{Eq.18}
\left.u_{I}^{+}\right|_{t=t_{0}}=\left.u_{F}^{+}\right|_{t=t_{0}}+\left.u_{F}^{-}\right|_{t=t_{0}},
\end{equation}
\begin{equation}\label{Eq.19}
\left.\frac{\partial u_{I}^{+}}{\partial t}\right|_{t=t_{0}}=\left.\frac{\partial u_{F}^{+}}{\partial t}\right|_{t=t_{0}}+\left.\frac{\partial u_{F}^{-}}{\partial t}\right|_{t=t_{0}}.
\end{equation}

For the time interface at $t=t_1$, using continuity conditions, we obtain:
\begin{equation}\label{Eq.20}
\left.u_{F}^{+}\right|_{t=t_{1}}+\left.u_{F}^{-}\right|_{t=t_{1}}=\left.u_{O}^{+}\right|_{t=t_{1}}+\left.u_{O}^{-}\right|_{t=t_{1}},
\end{equation}
\begin{equation}\label{Eq.21}
\left.\frac{\partial u_{F}^{+}}{\partial t}\right|_{t=t_{1}}+\left.\frac{\partial u_{F}^{-}}{\partial t}\right|_{t=t_{1}}=\left.\frac{\partial u_{O}^{+}}{\partial t}\right|_{t=t_{1}}+\left.\frac{\partial u_{O}^{-}}{\partial t}\right|_{t=t_{1}}.
\end{equation}

To simplify the derivations, we define:
\begin{equation}\label{Eq.22}
\kappa_{n}=\kappa_{0}+n\kappa_{m},
\end{equation}
\begin{equation}\label{Eq.23}
\hat{\omega}_{s,n}^{+/-}=\hat{\omega}_{s}^{+/-}+n\omega_{m}.
\end{equation}

Substituting \textcolor{blue}{Eqs. (\ref{Eq.12})-(\ref{Eq.14})} into \textcolor{blue}{Eqs. (\ref{Eq.18})-(\ref{Eq.19})}, we have:
\begin{equation}\label{Eq.24}
A_{0}e^{-i\kappa_{0}x}=\sum_{s=-N}^{N}\left(C_{s}\sum_{n=-N}^{N}\hat{u}_{s,n}^{+}e^{-i\kappa_{n}x}+D_{s}\sum_{n=-N}^{N}\hat{u}_{s,n}^{-}e^{-i\kappa_{n}x}\right),
\end{equation}
\begin{equation}\label{Eq.25}
\omega_{0}A_{0}e^{-i\kappa_{0}x}=\sum_{s=-N}^{N}\left(C_{s}\sum_{n=-N}^{N}\hat{u}_{s,n}^{+}\hat{\omega}_{s,n}^{+}e^{-i\kappa_{n}x}+D_{s}\sum_{n=-N}^{N}\hat{u}_{s,n}^{-}\hat{\omega}_{s,n}^{-}e^{-i\kappa_{n}x}\right).
\end{equation}

Similarly, substituting \textcolor{blue}{Eqs. (\ref{Eq.13})-(\ref{Eq.16})} into \textcolor{blue}{Eqs. (\ref{Eq.20})-(\ref{Eq.21})}, yields:
\begin{equation}\label{Eq.26}
\sum_{s=-N}^{N}\left[C_{s}\sum_{n=-N}^{N}\hat{u}_{s,n}^{+}e^{i(\hat{\omega}_{s,n}^{+}\Delta t-\kappa_{n}x)}+D_{s}\sum_{n=-N}^{N}\hat{u}_{s,n}^{-}e^{i(\hat{\omega}_{s,n}^{-}\Delta t-\kappa_{n}x)}\right]=\sum_{n=-N}^{N}\left[\tilde{T}_{n}e^{i(\tilde{\omega}_{n}^{+}\Delta t-\kappa_{n}x)}+\tilde{R}_{n}e^{i(\tilde{\omega}_{n}^{-}\Delta t-\kappa_{n}x)}\right],
\end{equation}
\begin{equation}\label{Eq.27}
\sum_{s=-N}^{N}\left[C_{s}\sum_{n=-N}^{N}\hat{u}_{s,n}^{+}\hat{\omega}_{s,n}^{+}e^{i(\hat{\omega}_{s,n}^{+}\Delta t-\kappa_{n}x)}+D_{s}\sum_{n=-N}^{N}\hat{u}_{s,n}^{-}\hat{\omega}_{s,n}^{-}e^{i(\hat{\omega}_{s,n}^{-}\Delta t-\kappa_{n}x)}\right]=\sum_{n=-N}^{N}\left[\tilde{T}_{n}\tilde{\omega}_{n}^{+}e^{i(\tilde{\omega}_{n}^{+}\Delta t-\kappa_{n}x)}+\tilde{R}_{n}\tilde{\omega}_{n}^{-}e^{i(\tilde{\omega}_{n}^{-}\Delta t-\kappa_{n}x)}\right].
\end{equation}

Note that, the spatial components of different modes are orthogonal, namely:
\begin{equation}\label{Eq.28}
\frac{1}{\lambda_{m}}\int_{-\lambda_{m}/2}^{\lambda_{m}/2}\psi_{p}(x)\cdot\psi_{q}^{*}(x)dx=
\begin{cases}
1&p=q\\
0&p\neq q
\end{cases},
\end{equation}
where, $\psi_{p}(x)=e^{-i\kappa_{p}x}$, $\psi_{q}(x)=e^{-i\kappa_{q}x}$ and the superscript $^*$ represents complex conjugate.

Hence, multiplying both sides of \textcolor{blue}{Eqs. (\ref{Eq.24})-(\ref{Eq.25})} by $e^{i\kappa_{-N}x}$, $\cdots$, $e^{i\kappa_{0}x}$, $\cdots$, $e^{i\kappa_{N}x}$ in sequence, applying the orthogonal relation shown in \textcolor{blue}{Eq. (\ref{Eq.28})}, we obtain:
\begin{equation}\label{Eq.29}
\left.
\begin{array}{lr}
0=\sum\limits_{s=-N}\limits^{N}C_{s}\hat{u}_{s,-N}^{+}+\sum\limits_{s=-N}\limits^{N}D_{s}\hat{u}_{s,-N}^{-},\\
\vdots\\
A_{0}=\sum\limits_{s=-N}\limits^{N}C_{s}\hat{u}_{s,0}^{+}+\sum\limits_{s=-N}\limits^{N}D_{s}\hat{u}_{s,0}^{-},\\
\vdots\\
0=\sum\limits_{s=-N}\limits^{N}C_{s}\hat{u}_{s,N}^{+}+\sum\limits_{s=-N}\limits^{N}D_{s}\hat{u}_{s,N}^{-},\\
\end{array}
\right\}(2N+1)
\end{equation}
\begin{equation}\label{Eq.30}
\left.
\begin{array}{lr}
0=\sum\limits_{s=-N}\limits^{N}C_{s}\hat{u}_{s,-N}^{+}\hat{\omega}_{s,-N}^{+}+\sum\limits_{s=-N}\limits^{N}D_{s}\hat{u}_{s,-N}^{-}\hat{\omega}_{s,-N}^{-},\\
\vdots\\
\omega_{0}A_{0}=\sum\limits_{s=-N}\limits^{N}C_{s}\hat{u}_{s,0}^{+}\hat{\omega}_{s,0}^{+}+\sum\limits_{s=-N}\limits^{N}D_{s}\hat{u}_{s,0}^{-}\hat{\omega}_{s,0}^{-},\\
\vdots\\
0=\sum\limits_{s=-N}\limits^{N}C_{s}\hat{u}_{s,N}^{+}\hat{\omega}_{s,N}^{+}+\sum\limits_{s=-N}\limits^{N}D_{s}\hat{u}_{s,N}^{-}\hat{\omega}_{s,N}^{-}.
\end{array}
\right\}(2N+1)
\end{equation}

In the same way, by multiplying both sides of \textcolor{blue}{Eqs. (\ref{Eq.26})-(\ref{Eq.27})} by $e^{i\kappa_{-N}x}$, $\cdots$, $e^{i\kappa_{0}x}$, $\cdots$, $e^{i\kappa_{N}x}$ in sequence and applying the orthogonal relation shown in \textcolor{blue}{Eq. (\ref{Eq.28})}, yields:
\begin{equation}\label{Eq.31}
\left.
\begin{array}{lr}
\sum\limits_{s=-N}\limits^{N}C_{s}\hat{u}_{s,-N}^{+}e^{i\hat{\omega}_{s,-N}^{+}\Delta t}+\sum\limits_{s=-N}\limits^{N}D_{s}\hat{u}_{s,-N}^{-}e^{i\hat{\omega}_{s,-N}^{-}\Delta t}=\tilde{T}_{-N}e^{i\tilde{\omega}_{-N}^{+}\Delta t}+\tilde{R}_{-N}e^{i\tilde{\omega}_{-N}^{-}\Delta t},\\
\vdots\\
\sum\limits_{s=-N}\limits^{N}C_{s}\hat{u}_{s,0}^{+}e^{i\hat{\omega}_{s,0}^{+}\Delta t}+\sum\limits_{s=-N}\limits^{N}D_{s}\hat{u}_{s,0}^{-}e^{i\hat{\omega}_{s,0}^{-}\Delta t}=\tilde{T}_{0}e^{i\tilde{\omega}_{0}^{+}\Delta t}+\tilde{R}_{0}e^{i\tilde{\omega}_{0}^{-}\Delta t},\\
\vdots\\
\sum\limits_{s=-N}\limits^{N}C_{s}\hat{u}_{s,N}^{+}e^{i\hat{\omega}_{s,N}^{+}\Delta t}+\sum\limits_{s=-N}\limits^{N}D_{s}\hat{u}_{s,N}^{-}e^{i\hat{\omega}_{s,N}^{-}\Delta t}=\tilde{T}_{N}e^{i\tilde{\omega}_{N}^{+}\Delta t}+\tilde{R}_{N}e^{i\tilde{\omega}_{N}^{-}\Delta t},\\
\end{array}
\right\}(2N+1)
\end{equation}
\begin{equation}\label{Eq.32}
\left.
\begin{array}{lr}
\sum\limits_{s=-N}\limits^{N}C_{s}\hat{u}_{s,-N}^{+}\hat{\omega}_{s,-N}^{+}e^{i\hat{\omega}_{s,-N}^{+}\Delta t}+\sum\limits_{s=-N}\limits^{N}D_{s}\hat{u}_{s,-N}^{-}\hat{\omega}_{s,-N}^{-}e^{i\hat{\omega}_{s,-N}^{-}\Delta t}=\tilde{T}_{-N}\tilde{\omega}_{-N}^{+}e^{i\tilde{\omega}_{-N}^{+}\Delta t}+\tilde{R}_{-N}\tilde{\omega}_{-N}^{-}e^{i\tilde{\omega}_{-N}^{-}\Delta t},\\
\vdots\\
\sum\limits_{s=-N}\limits^{N}C_{s}\hat{u}_{s,0}^{+}\hat{\omega}_{s,0}^{+}e^{i\hat{\omega}_{s,0}^{+}\Delta t}+\sum\limits_{s=-N}\limits^{N}D_{s}\hat{u}_{s,0}^{-}\hat{\omega}_{s,0}^{-}e^{i\hat{\omega}_{s,0}^{-}\Delta t}=\tilde{T}_{0}\tilde{\omega}_{0}^{+}e^{i\tilde{\omega}_{0}^{+}\Delta t}+\tilde{R}_{0}\tilde{\omega}_{0}^{-}e^{i\tilde{\omega}_{0}^{-}\Delta t},\\
\vdots\\
\sum\limits_{s=-N}\limits^{N}C_{s}\hat{u}_{s,N}^{+}\hat{\omega}_{s,N}^{+}e^{i\hat{\omega}_{s,N}^{+}\Delta t}+\sum\limits_{s=-N}\limits^{N}D_{s}\hat{u}_{s,N}^{-}\hat{\omega}_{s,N}^{-}e^{i\hat{\omega}_{s,N}^{-}\Delta t}=\tilde{T}_{N}\tilde{\omega}_{N}^{+}e^{i\tilde{\omega}_{N}^{+}\Delta t}+\tilde{R}_{N}\tilde{\omega}_{N}^{-}e^{i\tilde{\omega}_{N}^{-}\Delta t}.
\end{array}
\right\}(2N+1)
\end{equation}

\textcolor{blue}{Eqs. (\ref{Eq.29})-(\ref{Eq.30})} and \textcolor{blue}{Eqs. (\ref{Eq.31})-(\ref{Eq.32})} can be rewritten in matrix form, resulting in the following expressions, respectively:
\begin{equation}\label{Eq.33}
\mathbf{M_1}A_0=\mathbf{M_2}\begin{bmatrix}C_{-N}&\cdots&C_{N}&D_{-N}&\cdots&D_{N}\end{bmatrix}^{T},
\end{equation}
\begin{equation}\label{Eq.34}
\mathbf{M_3}\begin{bmatrix}C_{-N}&\cdots&C_{N}&D_{-N}&\cdots&D_{N}\end{bmatrix}^{T}=\mathbf{M_4}\begin{bmatrix}\tilde{T}_{-N}&\cdots&\tilde{T}_{N}&\tilde{R}_{-N}&\cdots&\tilde{R}_{N}\end{bmatrix}^{T}.
\end{equation}

From \textcolor{blue}{Eqs. (\ref{Eq.33})-(\ref{Eq.34})}, we ultimately obtain the following scattering relation in matrix form:
\begin{equation}\label{Eq.35}
\begin{bmatrix}\tilde{T}_{-N}&\cdots&\tilde{T}_{N}&\tilde{R}_{-N}&\cdots&\tilde{R}_{N}\end{bmatrix}^{T}=\mathbf{S}A_{0},
\end{equation}
where $\mathbf{S}=(\mathbf{M_4})^{-1}\mathbf{M_3}(\mathbf{M_2})^{-1}\mathbf{M_1}$. A detailed description of the transfer matrices $\mathbf{M_1}$, $\mathbf{M_2}$, $\mathbf{M_3}$ and $\mathbf{M_4}$ is presented in \textcolor{blue}{Appendix \ref{App.C}}. At this stage, we can determine the transmission and reflection coefficients for the $n^\mathrm{th}$-order scattering mode as:
\begin{equation}\label{Eq.36}
T_{n}=\left|\frac{\tilde{T}_{n}}{A_{0}}\right|=\left|\mathbf{S}_{N+1+n,1}\right|,
\end{equation}
\begin{equation}\label{Eq.37}
R_{n}=\left|\frac{\tilde{R}_{n}}{A_{0}}\right|=\left|\mathbf{S}_{3N+2+n,1}\right|.
\end{equation}
 
The above theoretical model applies to both subsonic and supersonic modulation regimes. In \textcolor{blue}{Sections \ref{sec3}} and \textcolor{blue}{\ref{sec4}}, we investigate the scattering behavior under subsonic and supersonic modulation, respectively, based on the mode-coupling theory developed.

\section{Temporal scattering behavior for subsonic spatiotemporal modulation}
\label{sec3}
We analyze the temporal scattering behaviors associated with a subsonic modulation of the temporal interlayer using the presented mode-coupling theory. To this purpose, we select the modulation parameters as follows: $V=0.2$, $\alpha_m=0.1$, $\kappa_m=10\;\mathrm{rad/m}$, $\omega_m=2\;\mathrm{rad/s}$, and $\Delta t=4\pi$.

\subsection{Temporal transmission and reflection coefficients under subsonic modulation}
\label{subsec3.1}
\begin{figure}[h]   \centering
\includegraphics[width=1\linewidth]{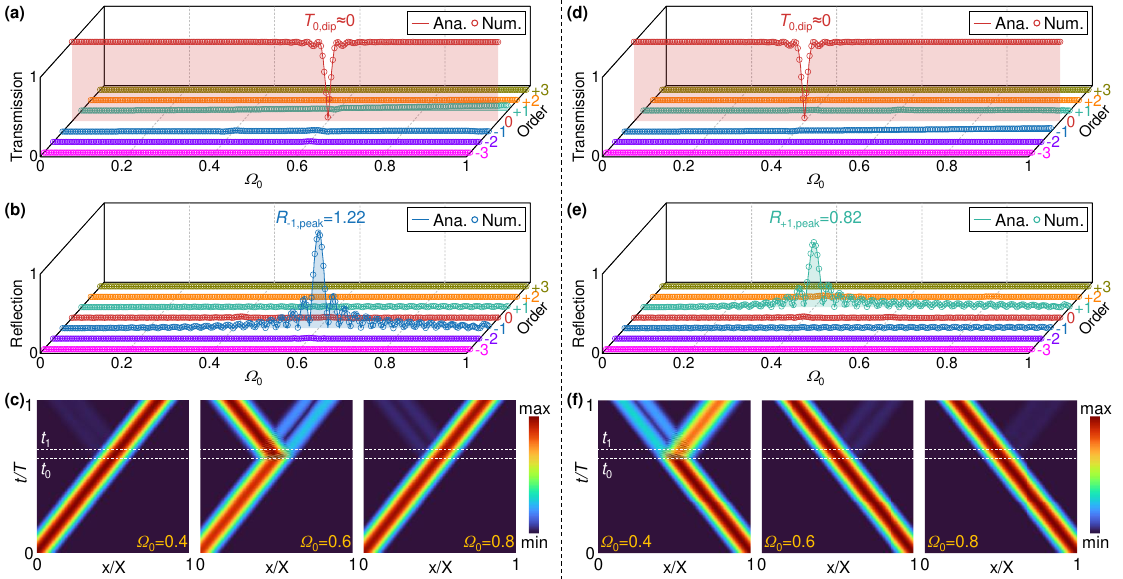}
\caption{Temporal scattering coefficients and wavefields for subsonic modulation, with $V=0.2$ and $\alpha_m = 0.1$. (a)-(c) Positive incidence. Analytical and numerical (a) transmission and (b) reflection coefficients. (c) Time-evolving wavefields at different incident frequencies $\mathit{\Omega}_0 = 0.4$, $0.6$, and $0.8$. (d)-(f) Negative incidence. Analytical and numerical (d) transmission and (e) reflection coefficients. (f) Time-evolving wavefields at different incident frequencies $\mathit{\Omega}_0 = 0.4$, $0.6$, and $0.8$, respectively.}
\label{Fig.4}
\end{figure}

We begin by considering the case of positive incidence. Analytical results are compared with numerical ones obtained from finite difference time domain (FDTD) simulations (see \textcolor{blue}{Appendix \ref{App.D}} for more simulation details). The analytical and numerical transmission and reflection coefficients for different incident frequencies are presented in \textcolor{blue}{Figs. \ref{Fig.4}}(a) and (b). As expected, in the frequency range $\mathit{\Omega}_0 = [0, 1]$, the transmission coefficient associated with the fundamental, $0^\mathrm{th}$-order mode, is equal to 1 at most frequencies, with all other scattering coefficients being almost null. As the incident frequency approaches $\mathit{\Omega}_0=0.6$,  which corresponds to the forward directional $\mathit{\Omega}$-bandgap, the $0^\mathrm{th}$-order transmission coefficient drops to $T_{0,dip}\approx0$, whereas the $-1^\mathrm{st}$-order reflection coefficient gradually increases to a maximum value of $R_{-1,peak}=1.22$. Therefore, for positive incident waves, a subsonic modulation induces wave conversion between the $0^\mathrm{th}$-order transmission and the $-1^\mathrm{st}$-order reflection, accompanied by a frequency down-conversion $\tilde{\omega}_{-1}=\omega_0-\omega_m$ ($\tilde{\mathit{\Omega}}_{-1}=\mathit{\Omega}_0-V$). To confirm these observations, snapshots of temporal scattering wavefields obtained from FDTD simulations at different excitation frequencies, i.e., $\mathit{\Omega}_0 = 0.4$, 0.6, and 0.8, are shown in \textcolor{blue}{Fig. \ref{Fig.4}}(c). As expected, for $\mathit{\Omega}_0 = 0.4$ and $0.8$, the incident waves are completely transmitted. In contrast, for $\mathit{\Omega}_0 = 0.6$, the incident waves are significantly reflected with almost no transmission.

Let us now examine the temporal scattering behavior for the negative incidence case. The corresponding scattering coefficients for different incident frequencies are presented in \textcolor{blue}{Figs. \ref{Fig.4}}(d) and (e). For negative incidence, the $0^\mathrm{th}$-order transmission suddenly drops to $T_{0,dip} \approx 0$, while the $+1^\mathrm{st}$-order reflection rises to a maximum value of $R_{+1,peak} = 0.82$ within the backward directional $\mathit{\Omega}$-bandgap at $\mathit{\Omega}_0 = 0.4$. As a result, a conversion between the $0^\mathrm{th}$-order transmission and the $+1^\mathrm{st}$-order reflection is observed, accompanied by frequency up-conversion, $\tilde{\omega}_{+1} = \omega_0 + \omega_m$ ($\tilde{\mathit{\Omega}}_{+1} = \mathit{\Omega}_0 + V$). Snapshots of temporal scattering wavefields for different excitation frequencies are also provided, as shown in \textcolor{blue}{Fig. \ref{Fig.4}}(f). In this case, the incident wave centered at $\mathit{\Omega}_0 = 0.4$ is reflected, while the incident waves centered at $\mathit{\Omega}_0 = 0.6$ and 0.8 are fully transmitted.

\subsection{Nonreciprocal wave conversion induced by subsonic modulation}
\label{subsec3.2}
\begin{figure}[h!]   \centering
\includegraphics[width=1\linewidth]{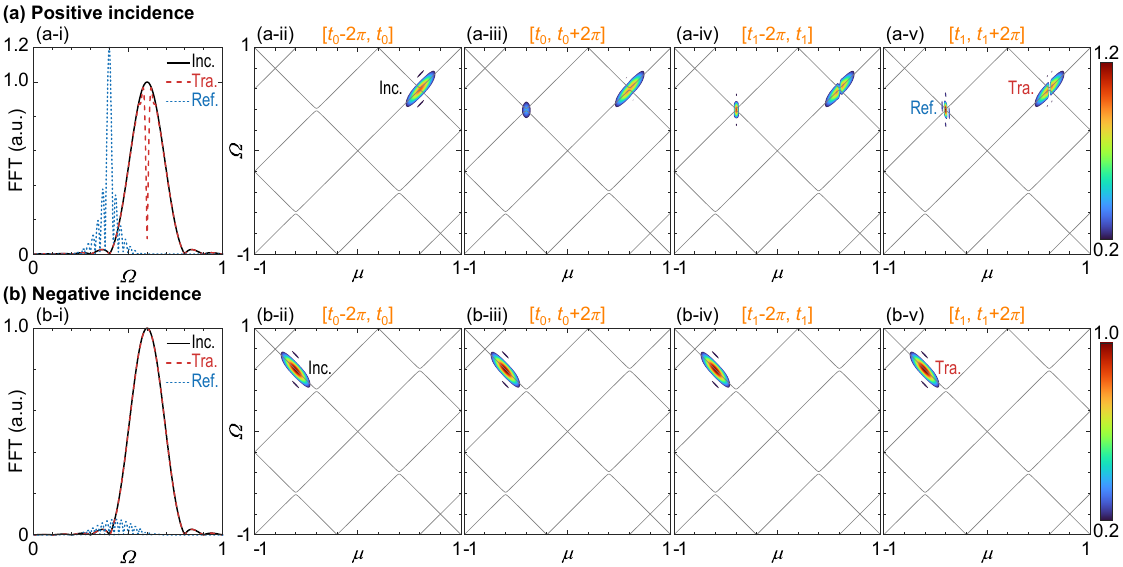}
\caption{Numerical demonstration of nonreciprocal wave conversion and frequency conversion under subsonic modulation. Normalized FFTs of the incident, transmitted, and reflected waves for (a-i) positive incidence and (b-i) negative incidence. Normalized 2D-FFTs of the incident domain $[t_0-2\pi, t_0]$, modulated domains $[t_0, t_0+2\pi]$ and $[t_1-2\pi, t_1]$, and scattering domain $[t_1, t_1+2\pi]$ for (a-ii)-(a-v) positive incidence and (b-ii)-(b-v) negative incidence. The background gray lines in (a-ii)-(a-v) and (b-ii)-(b-v) represent the corresponding dispersion diagram under subsonic modulation.}
\label{Fig.5}
\end{figure}

To visualize the energy transport and frequency conversion of waves passing through a spatiotemporal modulated waveguide with time interfaces, we evaluate the wavefield evolution in the frequency-wavenumber domain under a broadband excitation. For simplicity,  we focus only on the forward $\mathit{\Omega}$-bandgap located at $\mathit{\Omega}=0.6$. First, the frequency contents of the incident and scattering waves are evaluated by probing different regions of the time-evolving wavefield. The corresponding normalized FFT results of the incident, transmitted, and reflected waves for positive and negative incident excitations are presented in \textcolor{blue}{Figs. \ref{Fig.5}}(a-i) and (b-i). From \textcolor{blue}{Fig. \ref{Fig.5}}(a-i), we observe that the transmitted amplitude at the central frequency of bandgap, i.e. $\mathit{\Omega}_t=0.6$, is severely attenuated, while the amplitudes of other frequency components remain consistent with the incident wave. On the other hand, the reflected spectrum exhibits a pronounced peak at a distinct frequency from that of the incident wave, indicating wave conversion between the transmitted wave and a frequency-shifted reflected wave. In contrast, for the negative incidence case shown in \textcolor{blue}{Fig. \ref{Fig.5}}(b-i), the transmission spectrum closely matches the incident spectrum, while the magnitude of the reflection spectrum is negligible.

Next, we represent the signals in frequency-wavenumber domains through both spatial and temporal FFT (2D-FFT). The frequency and wavenumber distribution contours for the incident domain $[t_0-2\pi, t_0]$, modulated domains $[t_0, t_0+2\pi]$ and $[t_1-2\pi, t_1]$, and scattering domain $[t_1, t_1+2\pi]$ are presented in \textcolor{blue}{Figs. \ref{Fig.5}}(a-ii)-(a-v) and \textcolor{blue}{Figs. \ref{Fig.5}}(b-ii)-(b-v), for positive and negative incidence, respectively. 
To aid interpretation, the dispersion diagram of the subsonically modulated medium is superimposed as background gray lines. For the positive incidence case, the directional frequency bandgap associated with the positive subbranch is activated; consequently, the central energy of the incident wave is progressively redirected toward the negative subbranch. \textcolor{blue}{Figs.~\ref{Fig.5}}(a-ii)-(a-v) clearly demonstrate how an incident broadband wave centered at $\mathit{\Omega}_0 = 0.6$ is converted into a reflected narrowband wave centered at $\mathit{\Omega}_r = 0.4$ via interaction with the temporally modulated interlayer. In contrast, for the negative incidence case shown in \textcolor{blue}{Figs. \ref{Fig.5}}(b-ii)-(b-v), the excitation falls within the passband of the negative subbranch, resulting in a full transmission without frequency conversion.

\subsection{Parametric analysis for subsonic modulation}
\label{subsec3.3}

\begin{figure}[h!]   \centering
\includegraphics[width=0.75\linewidth]{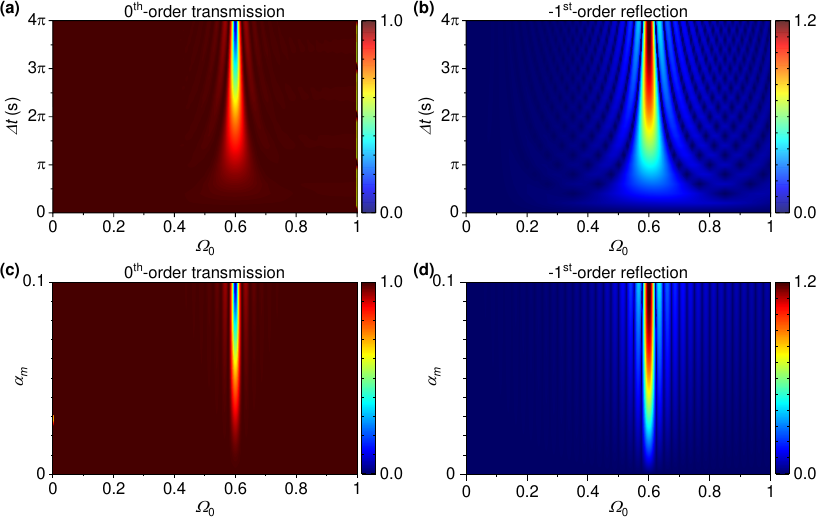}
\caption{Parametric analysis of modulation duration $\Delta t$ and amplitude $\alpha_m$ for positive incidence under subsonic modulation. (a) $0^\mathrm{th}$-order transmission and (b) $-1^\mathrm{st}$-order reflection coefficient profiles varying as modulation duration $\Delta t$. (c) $0^\mathrm{th}$-order transmission and (d) $-1^\mathrm{st}$-order reflection coefficient profiles varying as modulation amplitude $\alpha_m$.}
\label{Fig.6}
\end{figure}

We investigate the effect of modulation duration $\Delta t$ and amplitude $\alpha_m$ on scattering coefficients for subsonic modulation. As an illustration, we consider the positive incidence case. Based on the above analysis, wave energy is mainly transported between the $0^\mathrm{th}$-order transmission and the $-1^\mathrm{st}$-order reflection modes. For other orders, the scattered amplitudes can be neglected. Thus, here we focus on the variation of the $0^\mathrm{th}$-order transmission and the $-1^\mathrm{st}$-order reflection coefficients with modulation parameters. The coefficient contours of the dominant transmission and reflection modes versus the incident frequency and modulation duration are shown in \textcolor{blue}{Figs. \ref{Fig.6}}(a) and (b).
As expected, the transmission within the $\mathit{\Omega}$-bandgap decreases as the modulation duration increases, whereas the reflection within the $\mathit{\Omega}$-bandgap correspondingly increases. The influence of modulation amplitude on the dominant scattering profiles is illustrated in \textcolor{blue}{Figs.~\ref{Fig.6}}(c) and (d). It is evident that the transmission coefficient within the $\mathit{\Omega}$-bandgap decreases, while the reflection coefficient increases with increasing modulation amplitude. Therefore, the wave conversion effect becomes more pronounced as both the modulation duration $\Delta t$ and amplitude $\alpha_m$ increase.

\section{Temporal scattering behavior for supersonic spatiotemporal modulation}
\label{sec4}
We discuss the temporal scattering behavior induced by supersonic spatiotemporal modulation. Here, the modulation parameters are set to $V=2$, $\alpha_m=0.1$, $\kappa_m=10\;\mathrm{rad/m}$, $\omega_m=20\;\mathrm{rad/s}$, and $\Delta t=4\pi$, while the normalized frequency range of investigation is set as $\mathit{\Omega}_0=[0, 2]$.

\subsection{Temporal transmission and reflection coefficients under supersonic modulation}
\label{subsec4.1}

\begin{figure}[h]   \centering
\includegraphics[width=1\linewidth]{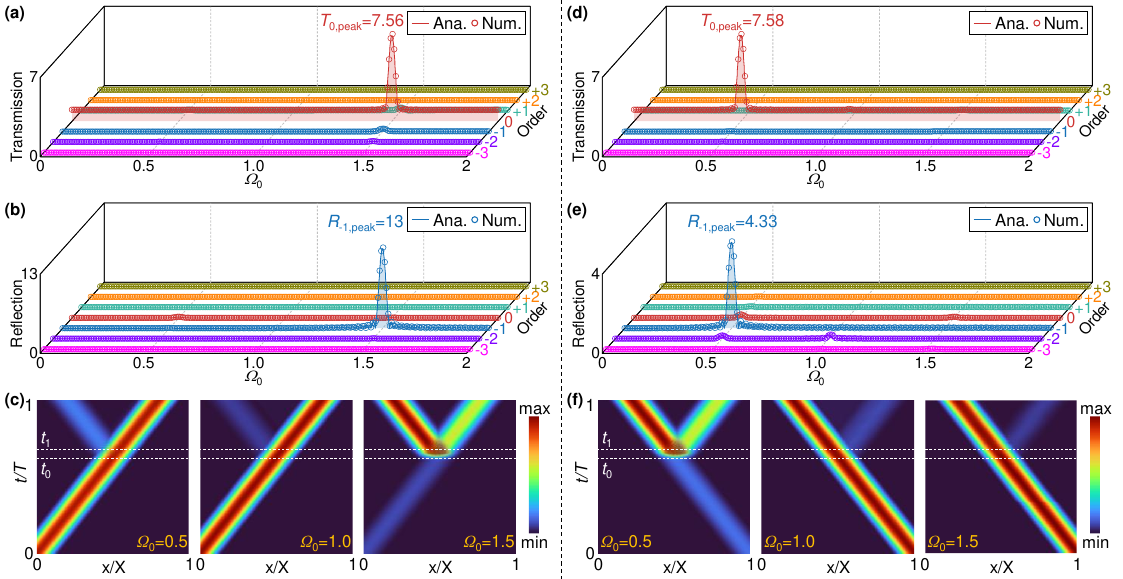}
\caption{Temporal scattering coefficients and wavefields for supersonic modulation, with $V=2$ and $\alpha_m = 0.1$. (a)-(c) Positive incidence. Analytical and numerical (a) transmission and (b) reflection coefficients. (c) Time-evolving wavefields at different incident frequencies $\mathit{\Omega}_0 = 0.5$, $1.0$, and $1.5$. (d)-(f) Negative incidence. Analytical and numerical (d) transmission and (e) reflection coefficients. (f) Time-evolving wavefields at different incident frequencies $\mathit{\Omega}_0 = 0.5$, $1.0$, and $1.5$, respectively.}
\label{Fig.7}
\end{figure}

First, we investigate the positive incidence case. The analytical and numerical scattering coefficients for varying incident frequencies are shown in \textcolor{blue}{Figs. \ref{Fig.7}}(a) and (b). As observed in the subsonic modulation scenario, outside the bandgap, the $0^\mathrm{th}$-order transmission dominates all scattering modes with a unitary coefficient. However, within the forward $\mu$-bandgap at $\mathit{\Omega}_0 = 1.5$, both the $0^\mathrm{th}$-order transmission and the $-1^\mathrm{st}$-order reflection coefficients increase rapidly, reaching peak values of $T_{0,peak} = 7.56$ and $R_{-1,peak} = 13$, respectively. This indicates that under supersonic modulation, unlike in the subsonic case, the pumping effect in the $\mu$-bandgap manifests as parametric amplification through both transmission and reflection, along with frequency down-conversion by reflection, $\tilde{\omega}_{-1} = \omega_0 - \omega_m$ ($\tilde{\mathit{\Omega}}_{-1} = \mathit{\Omega}_0 - V$). To corroborate these observations, we present snapshots of scattering wavefields for different excitation frequencies, i.e., $\mathit{\Omega}_0 = 0.5$, 1.0, and 1.5, as shown in \textcolor{blue}{Fig. \ref{Fig.7}}(c). As expected, the incident wave is simply transmitted without parametric amplification for $\mathit{\Omega}_0 = 0.5$ and 1.0. In contrast, for $\mathit{\Omega}_0 = 1.5$, both the transmitted and reflected waves exhibit parametric amplification with amplitudes significantly higher than that of the incident waves.

For the negative incidence scenario, we observe parametric amplification of the $0^\mathrm{th}$-order transmission and the $-1^\mathrm{st}$-order reflection at $\mathit{\Omega}_0 = 0.5$, with peak coefficients of $T_{0,peak} = 7.58$ and $R_{-1,peak} = 4.33$, as shown in \textcolor{blue}{Figs. \ref{Fig.7}}(d) and (e). Again, we confirm the predicted temporal scattering behaviors by providing time-evolving wavefields for different incident frequencies, as shown in \textcolor{blue}{Fig. \ref{Fig.7}}(f). A significant amplification of the temporally transmitted and reflected waves is observed within the backward $\mu$-bandgap at $\mathit{\Omega}_0 = 0.5$. A comparison of \textcolor{blue}{Figs. \ref{Fig.7}}(c) and (f) confirms the nonreciprocity of this parametric amplification effect.

\subsection{Nonreciprocal parametric amplification induced by supersonic modulation}
\begin{figure}[h]   \centering
\includegraphics[width=1\linewidth]{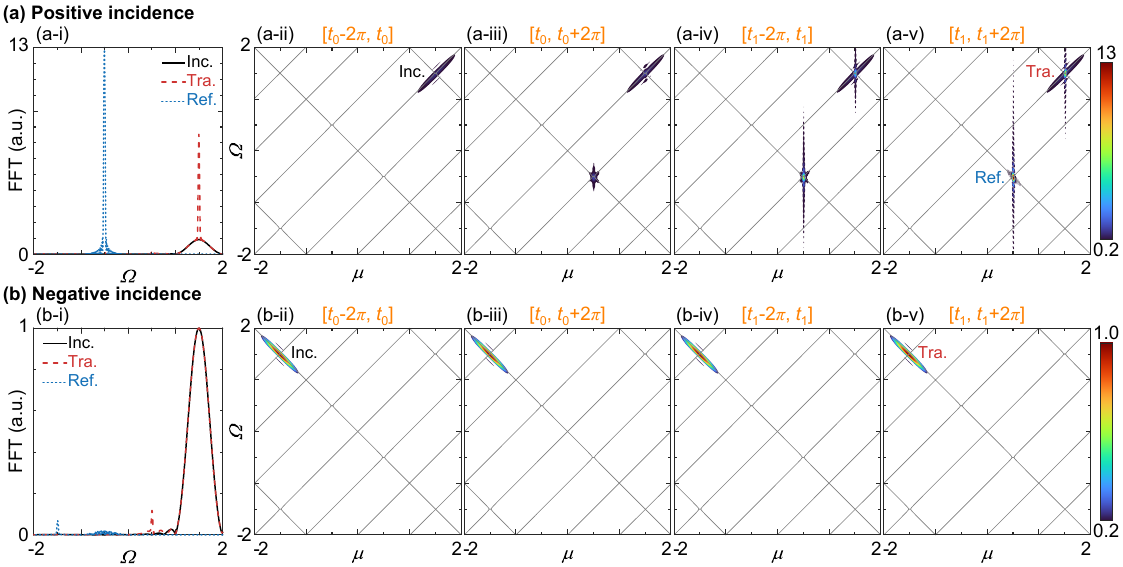}
\caption{Numerical demonstration of nonreciprocal parametric amplification and frequency conversion under supersonic modulation. Normalized FFTs of the incident, transmitted, and reflected waves for (a-i) positive incidence and (b-i) negative incidence. Normalized 2D-FFTs of the incident domain $[t_0-2\pi, t_0]$, modulated domains $[t_0, t_0+2\pi]$ and $[t_1-2\pi, t_1]$, and scattering domain $[t_1, t_1+2\pi]$ for (a-ii)-(a-v) positive incidence and (b-ii)-(b-v) negative incidence. The background gray lines in (a-ii)-(a-v) and (b-ii)-(b-v) represent the corresponding dispersion diagram under supersonic modulation.}
\label{Fig.8}
\end{figure}

As for the subsonic scenario, we analyze the frequency-wavenumber properties of wave packets injected from two opposite directions. As an illustration, we consider the case of a broadband input signal with a central frequency of $\mathit{\Omega}_0 = 1.5$, which corresponds to a forward directional $\mu$-bandgap. First, the normalized FFT results of the incident, transmitted, and reflected waves for both positive and negative excitations are presented in \textcolor{blue}{Figs.~\ref{Fig.8}}(a-i) and (b-i), respectively.
The transmitted amplitude at the central frequency of bandgap, i.e. $\mathit{\Omega}_t=1.5$, is considerably higher than the incident amplitude. Additionally, a reflection peak with a large magnitude is observed, though it occurs at a distinctly different central frequency, indicating frequency conversion via reflection. Conversely, for the negative incidence case shown in \textcolor{blue}{Fig.~\ref{Fig.8}}(b-i), the transmission spectrum matches the incident spectrum, and almost no reflection is observed, indicating complete transmission.

We corroborate these findings by analyzing the frequency-wavenumber content of the signals through 2D-FFTs performed on the incident domain $[t_0 - 2\pi,\, t_0]$, the modulated domains $[t_0,\, t_0 + 2\pi]$ and $[t_1 - 2\pi,\, t_1]$, and the scattering domain $[t_1,\, t_1 + 2\pi]$. For the positive incidence case, as shown in \textcolor{blue}{Figs.~\ref{Fig.8}}(a-ii)-(a-v), the supersonic modulation amplifies the wave energy of the transmitted and reflected waves. In addition, the reflected wave centered at $\mathit{\Omega}_r=-0.5$ occurs frequency conversion relative to the incident wave centered at $\mathit{\Omega}_0=1.5$. The dispersion diagram of the supersonically modulated medium is overlaid as background gray lines to aid interpretation. 

For reference, the results for negative excitation are presented in \textcolor{blue}{Figs.~\ref{Fig.8}}(b-i)-(b-v). In this case, the 2D-FFT contour of the transmitted wave is identical to that of the incident wave, indicating simple transmission without parametric amplification or frequency conversion.
 
\label{subsec4.2}

\subsection{Parametric analysis for supersonic modulation}
\label{subsec4.3}
\begin{figure}[h]   \centering
\includegraphics[width=0.75\linewidth]{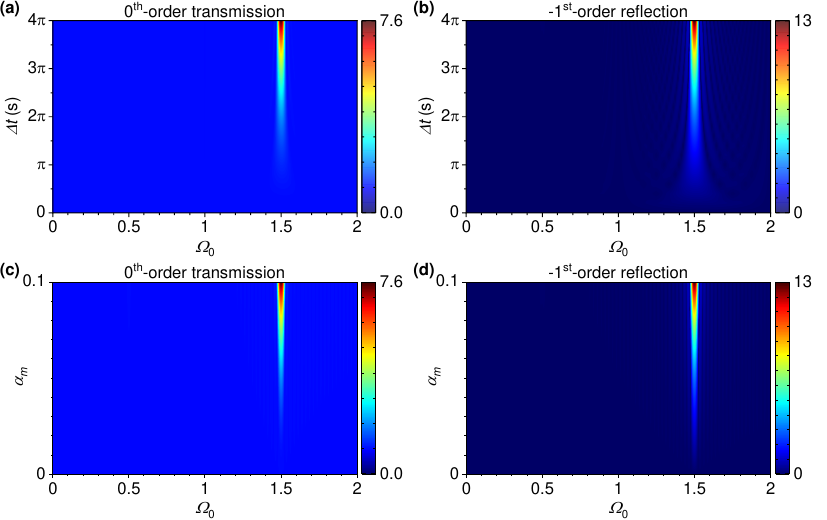}
\caption{Parametric analysis of modulation duration $\Delta t$ and amplitude $\alpha_m$ for positive incidence under supersonic modulation. (a) $0^\mathrm{th}$-order transmission and (b) $-1^\mathrm{st}$-order reflection coefficient profiles varying as modulation duration $\Delta t$. (c) $0^\mathrm{th}$-order transmission and (d) $-1^\mathrm{st}$-order reflection coefficient profiles varying as modulation amplitude $\alpha_m$.}
\label{Fig.9}
\end{figure}

We further study the effect of the modulation duration $\Delta t$ and amplitude $\alpha_m$ on the scattering coefficients for supersonic modulation. We consider positive incidence as an example.  Except for the dominant orders, i.e. the $0^\mathrm{th}$-order transmission and the $-1^\mathrm{st}$-order reflection, the scattered energy of other modes is almost null and thus can be ignored. Hence, we focus on the dominant scattering modes. The coefficient profiles of the $0^\mathrm{th}$-order transmission and the $-1^\mathrm{st}$-order reflection varying with the incident frequency and modulation duration are shown in \textcolor{blue}{Figs. \ref{Fig.9}}(a) and (b). The scattered energy outside the $\mu$-bandgap remains constant with modulation duration, whereas both the transmitted and reflected energy within the $\mu$-bandgap increase significantly with the increase of modulation duration. In addition, in \textcolor{blue}{Figs. \ref{Fig.9}}(c) and (d), we report the variations of the dominant scattering coefficients with the modulation amplitude. The dominant scattering coefficient outside the $\mu$-bandgap remains nearly constant with increasing modulation amplitude, while those within the $\mu$-bandgap increase dramatically as the modulation amplitude increases. Therefore, the amplification effect of both transmission and reflection within the $\mu$-bandgap can be enhanced by increasing the modulation duration $\Delta t$ and amplitude $\alpha_m$.

\section{Conclusions and prospects}
\label{sec5}
We presented a theoretical framework that enables analytical predictions of nonreciprocal scattering of elastic waves at time interfaces induced by spatiotemporal modulation. Within the modulated temporal interlayer, incident waves propagating from opposite directions are decomposed into distinct Floquet modes, which subsequently couple into multiple transmission and reflection orders in the scattering field. 

We analyzed two modulation regimes, each exhibiting unique dispersion characteristics and temporal scattering behaviors.  For subsonic modulation, we demonstrated nonreciprocal switching of transmission and reflection, along with frequency conversion via reflection. Under supersonic modulation, we observed nonreciprocal parametric amplification in both transmission and reflection, in addition to frequency conversion by reflection. These findings highlight the potential for controlled, time-growing wave amplification by tuning the parameters of the temporal interlayer modulation. Our results provide insights into the role of time interfaces in shaping elastic wave propagation within bounded spatiotemporally modulated systems.  

We anticipate that the proposed theoretical framework can be extended to other one-dimensional wave systems with a spatiotemporally modulated modulus. For example, the associated analysis is applicable to acoustic waves by replacing the elastic modulus $E(x,t)$ with a spatiotemporally modulated bulk modulus $K(x,t)$. A similar analogy applies to string waves, where $E(x,t)$ is replaced by a modulated tension $T(x,t)$, and the mass density is interpreted as the linear mass density.  

Future efforts will be devoted to extending our approach to account for material damping, which is inherently present in experimental realizations, by incorporating complex frequency modes associated with the dispersive properties of the modulated interlayer.

\section*{CRediT authorship contribution statement
}
\textbf{Yingrui Ye:} Conceptualization, Investigation, Formal analysis and modeling, Methodology, Visualization, Writing - original draft, Funding acquisition.
\textbf{Chunxia Liu:} Conceptualization, Investigation, Formal analysis and modeling, Methodology.
\textbf{Alessandro Marzani:} Conceptualization, Investigation, Formal analysis, Project administration, Supervision, Writing - review \& editing.
\textbf{Emanuele Riva:} Formal analysis, Supervision, Writing - review \& editing.
\textbf{Antonio Palermo:} Conceptualization, Investigation, Formal analysis, Project administration, Supervision, Writing - review \& editing, Funding acquisition.
\textbf{Xiaopeng Wang:} Conceptualization, Project administration, Supervision, Writing - review \& editing.

\section*{Declaration of competing interest}
The authors declare that they have no known competing financial interests or personal relationships that could have appeared to influence the work reported in this paper.

\section*{Acknowledgments}
A.P. acknowledges the funding received from the Italian Ministry of University and Research (MUR) for the "EXTREME" project (grant agreement 2022EZT2ZE, CUP: J53C24002870006). Y.Y. gratefully acknowledges the support from the China Scholarship Council (CSC Grant No. 202306280145).

\appendix
\section{Dispersion relation of spatiotemporal stiffness-modulated media}
\label{App.A}
\renewcommand{\thefigure}{A.\arabic{figure}}
\setcounter{figure}{0}
In the spatiotemporal modulated media with modulated elastic stiffness $E(x,t)$, due to the periodic nature in both space $x$ and time $t$ dimensions, $E(x,t)$ may be written as a Fourier series:
\begin{equation}\label{Eq.A1}\tag{A1}
E(x,t)=\sum_{p=-\infty}^{+\infty}\hat{E}_{p}e^{ip(\omega_m t-\kappa_m x)},
\end{equation}
where $\hat{E}_{p}$ is the Fourier coefficient given by:
\begin{equation}\label{Eq.S2}\tag{S2}
\hat{E}_{p}=\frac{1}{T_{m}}\frac{1}{\lambda_{m}}\int_{-T_m/2}^{T_m/2}\int_{-\lambda_m/2}^{\lambda_m/2}E(x,t)e^{-ip(\omega_{m}t-\kappa_{m}x)}dxdt.
\end{equation}

Hence, a complete solution to \textcolor{blue}{Eq. (\ref{Eq.2})} could be cast into the generalized Floquet form:
\begin{equation}\label{Eq.A3}\tag{A3}
u(x,t)=e^{i(\omega t-\kappa x)}\sum_{n=-\infty}^{+\infty}\hat{u}_{n}e^{in(\omega_m t-\kappa_m x)},
\end{equation}
where $\hat{u}_{n}$ is the amplitude of the $n^{\mathrm{th}}$-order Floquet-Bloch mode.

Substituting \textcolor{blue}{Eqs. (\ref{Eq.A1})} and \textcolor{blue}{(\ref{Eq.A3})} into \textcolor{blue}{Eq. (\ref{Eq.2})}, we have
\begin{equation}\label{Eq.A4}\tag{A4}
\begin{gathered}   
(\kappa+n\kappa_{m})(\kappa+p\kappa_{m})\sum_{n=-\infty}^{+\infty}\sum_{p=-\infty}^{+\infty}\hat{u}_{n}\hat{E}_{p-n}e^{ip(\omega_{m}t-\kappa_{m}x)}-\rho_0(\omega+n\omega_{m})^2\sum_{n=-\infty}^{+\infty}\hat{u}_{n}e^{in(\omega_{m}t-\kappa_{m}x)}=0.
\end{gathered}
\end{equation}

Multiplying both sides of \textcolor{blue}{Eq. (\ref{Eq.A4})} by $e^{-im(\omega_mt-\kappa_mx)}\quad(m=0,\pm1,\pm2,\cdots)$, and integrating over the spatial and temporal periods $[-T_m/2, T_m/2]$ and $[-\lambda_m/2, \lambda_m/2]$, \textcolor{blue}{Eq. (\ref{Eq.A4})} yields the following quadratic eigenvalue problem:
\begin{equation}\label{Eq.A5}\tag{A5}
\begin{gathered}
\sum_{n=-\infty}^{+\infty}\Bigl[(\kappa+m\kappa_{m})(\kappa+n\kappa_{m})\hat{E}_{m-n}-\rho_{0}(\omega+n\omega_{m})^{2}\delta_{mn}\Bigr]\hat{u}_{n}=0, m=0,\pm1,\pm2,\cdots,
\end{gathered}
\end{equation}
where $\delta_{mn}$ is the Kronecker-delta tensor. 

To obtain a nontrivial solution to \textcolor{blue}{Eq. (\ref{Eq.A5})}, the coefficient determinant of the amplitudes $\hat{u}_n$ should vanish. Thus, the dispersion relation for spatiotemporal stiffness-modulated media can be obtained as:
\begin{equation}\label{Eq.A6}\tag{A6}
\det
\begin{bmatrix}\cdots&\cdots&\cdots&\cdots&\cdots\\
\cdots&\beta_{-1,-1}-\zeta_{-1}&\beta_{-1,0}&\beta_{-1,1}&\cdots\\
\cdots&\beta_{0,-1}&\beta_{0,0}-\zeta_{0}&\beta_{0,1}&\cdots\\
\cdots&\beta_{1,-1}&\beta_{1,0}&\beta_{1,1}-\zeta_{1}&\cdots\\
\cdots&\cdots&\cdots&\cdots&\cdots\\
\end{bmatrix}=0,
\end{equation}
where $\beta_{m,n}=(\kappa+m\kappa_{m})(\kappa+n\kappa_{m})\hat{E}_{m-n}$, $\zeta_{n}=\rho_0(\omega+n\omega_{m})^2$. Considering a truncation order of $N$, i.e., $m,n=-N, \cdots, +N$, we can obtain $4N+2$ eigenvalues $\hat{\omega}_{s}^{+/-}$ and $4N+2$ eigenvectors $\mathbf{\hat{U}^{+/-}_s}=\left[\begin{matrix}{\hat{u}^{+/-}_{s,-N},}&{\cdots,}&{\hat{u}^{+/-}_{s,N}}\end{matrix}\right]^{T}$ ($s=-N, \cdots, +N$) from \textcolor{blue}{Eq. (\ref{Eq.A6})}.

\section{Evolution of the dispersion diagram with modulation velocity}
\label{App.B}
\renewcommand{\thefigure}{B.\arabic{figure}}
\setcounter{figure}{0}
As shown in \textcolor{blue}{Fig. \ref{Fig.B.1}}, we investigate the band structures of spatiotemporally modulated medium bounded by time interfaces with different modulation velocities by maintaining the modulation amplitude $\alpha_m=0.3$. According to \textcolor{blue}{Eq. (3)}, the wave velocity in the modulated medium falls within the range of $0.84c_0<c<1.14c_0$. As the modulation velocity increases, the modulation type gradually changes from subsonic modulation to hybrid modulation and finally to supersonic modulation.

We present the band diagram evolution with increasing modulation velocity, as shown in \textcolor{blue}{Figs. \ref{Fig.B.1}}(a)-(i). When $V<0.84$, the modulation velocity is always less than the wave velocity in the modulated medium, and subsonic modulation occurs. We can find that the interaction between different branches leads to frequency bandgaps, which are characterized by the absence of frequency solutions within the bandgaps, as shown in \textcolor{blue}{Figs. \ref{Fig.B.1}}(a)-(c). When $0.84<V<1.14$, the modulation velocity is between the minimum wave velocity and the maximum wave velocity, and hybrid modulation occurs. In this case, the frequency bandgaps gradually transform into wavenumber bandgaps, and these two types of bandgaps can coexist, as shown in \textcolor{blue}{Figs. \ref{Fig.B.1}}(d)-(f). One can see, especially from \textcolor{blue}{Fig. \ref{Fig.B.1}}(d), that hybrid modulation can lead to a tilted wavenumber bandgap, where the frequency has an imaginary part but its real part is not constant. When $V>1.14$, the modulation velocity is always greater than the wave velocity in the modulated medium, and supersonic modulation occurs. The frequency bandgaps are now completely converted into wavenumber bandgaps, characterized by a constant real part of frequency and a non-zero imaginary part of frequency within the bandgaps, as shown in \textcolor{blue}{Figs. \ref{Fig.B.1}}(g)-(i).

\begin{figure}[h]   \centering
\includegraphics[width=0.9\linewidth]{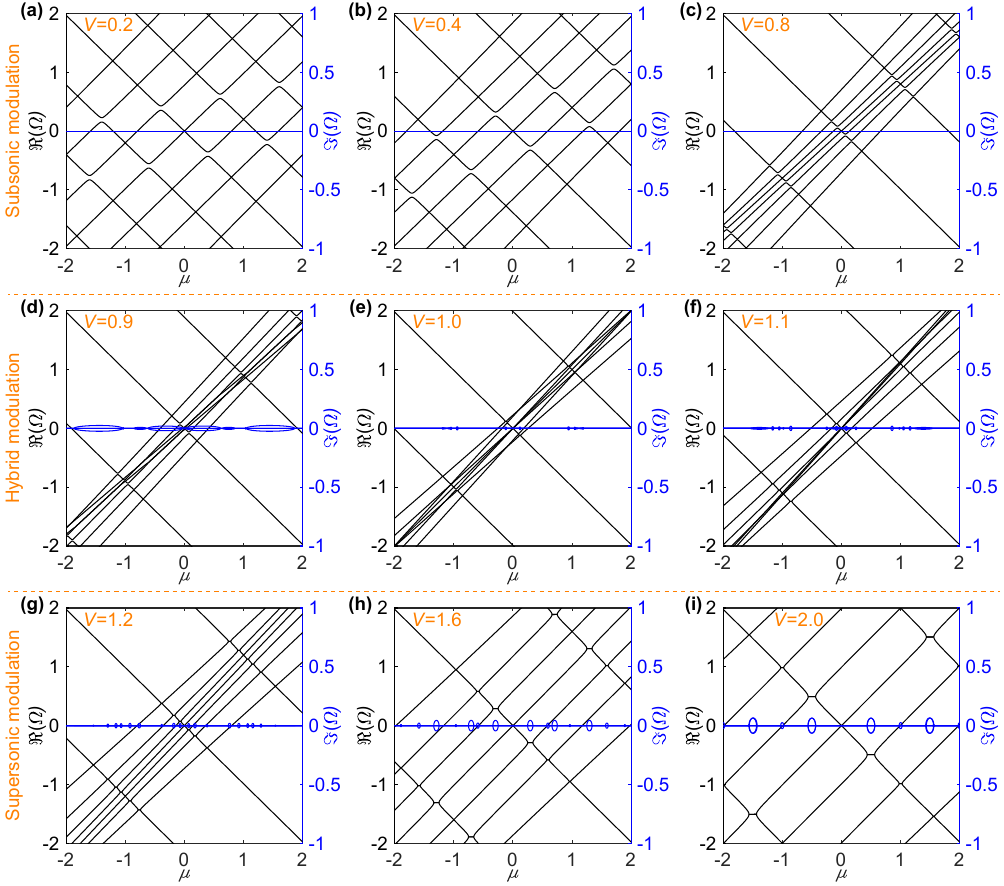}
\caption{Dispersion diagrams with the constant modulation amplitude $\alpha_m=0.3$ and different dimensionless modulation velocities (a) $V = 0.2$, (b) $V = 0.4$, (c) $V = 0.8$, (d) $V = 0.9$, (e) $V = 1.0$, (f) $V = 1.1$, (g) $V = 1.2$, (h) $V = 1.6$, and (i) $V = 2.0$, respectively.}
\label{Fig.B.1}
\end{figure}

In general, the relative magnitude of the modulation velocity and wave velocity in the modulated medium determines the interaction mechanism between different branches, resulting in frequency bandgaps or wavenumber bandgaps. Note that the bandgap transformation occurs in the hybrid modulation range, which manifests as a competition between the two interaction mechanisms. In the hybrid modulation range, small changes in the modulation velocity can lead to large differences in the bandgaps. Thereby, in the main text, we only investigate the scattering behavior under subsonic modulation and supersonic modulation. In addition, we notice that for both subsonic and supersonic modulation, the bandgap width becomes narrower as the modulation velocity approaches the hybrid modulation range. This means that to obtain a wider frequency bandgap, we should choose a smaller modulation velocity, while to obtain a wider wavenumber bandgap, we should choose a larger modulation velocity.

\section{Transfer matrices of mode-coupling theory for modeling temporal scattering behavior}
\label{App.C}
\renewcommand{\thefigure}{C.\arabic{figure}}
\setcounter{figure}{0}
$\mathbf{M_1}$ is a matrix of size $(4N+2)\times1$ and given by:
\begin{equation}\label{Eq.C1}\tag{C1}
\mathbf{M_{1}}=\begin{bmatrix}0&\cdots&1&\cdots&0&0&\cdots&\omega_{0}&\cdots&0\end{bmatrix}^{T}.
\end{equation}

$\mathbf{M_2}$ is a matrix of size $(4N+2)\times(4N+2)$ and given by:
\begin{equation}\label{Eq.C2}\tag{C2}
\mathbf{M_{2}}=
\begin{bmatrix}
{{\hat{u}_{-N,-N}^{+}}}&{\cdots}&{{\hat{u}_{N,-N}^{+}}}&{\hat{u}_{-N,-N}^{-}}&{\cdots}&{{\hat{u}_{N,-N}^{-}}}\\
{\vdots}&{\vdots}&{\vdots}&{\vdots}&{\vdots}&{\vdots}\\
{{\hat{u}_{-N,N}^{+}}}&{\cdots}&{{\hat{u}_{N,N}^{+}}}&{{\hat{u}_{-N,N}^{-}}}&{\cdots}&{{\hat{u}_{N,N}^{-}}}\\
{\hat{u}_{-N,-N}^{+}\hat{\omega}_{-N,-N}^{+}}&{\cdots}&{\hat{u}_{N,-N}^{+}\hat{\omega}_{N,-N}^{+}}&{\hat{u}_{-N,-N}^{-}\hat{\omega}_{-N,-N}^{-}}&{\cdots}&{\hat{u}_{N,-N}^{-}\hat{\omega}_{N,-N}^{-}}\\
{\vdots}&{\vdots}&{\vdots}&{\vdots}&{\vdots}&{\vdots}\\
{\hat{u}_{-N,N}^{+}\hat{\omega}_{-N,N}^{+}}&{\cdots}&{\hat{u}_{N,N}^{+}\hat{\omega}_{N,N}^{+}}&{\hat{u}_{-N,N}^{-}\hat{\omega}_{-N,N}^{-}}&{\cdots}&{\hat{u}_{N,N}^{-}\hat{\omega}_{N,N}^{-}}
\end{bmatrix}.
\end{equation}

$\mathbf{M_3}$ is a matrix of size $(4N+2)\times(4N+2)$ and given by:
\begin{footnotesize}
\begin{equation}\label{Eq.C3}\tag{C3}
\mathbf{M_{3}}=
\begin{bmatrix}
{\hat{u}_{-N,-N}^{+}e^{i\hat{\omega}_{-N,-N}^{+}\Delta t}}&{\cdots}&{\hat{u}_{N,-N}^{+}e^{i\hat{\omega}_{N,-N}^{+}\Delta t}}&\hat{u}_{-N,-N}^{-}e^{i\hat{\omega}_{-N,-N}^{-}\Delta t}&{\cdots}&{\hat{u}_{N,-N}^{-}e^{i\hat{\omega}_{N,-N}^{-}\Delta t}}\\
{\vdots}&{\vdots}&{\vdots}&{\vdots}&{\vdots}&{\vdots}\\
{\hat{u}_{-N,N}^{+}e^{i\hat{\omega}_{-N,N}^{+}\Delta t}}&{\cdots}&{\hat{u}_{N,N}^{+}e^{i\hat{\omega}_{N,N}^{+}\Delta t}}&{\hat{u}_{-N,N}^{-}e^{i\hat{\omega}_{-N,N}^{-}\Delta t}}&{\cdots}&{\hat{u}_{N,N}^{-}e^{i\hat{\omega}_{N,N}^{-}\Delta t}}\\
{\hat{u}_{-N,-N}^{+}\hat{\omega}_{-N,-N}^{+}e^{i\hat{\omega}_{-N,-N}^{+}\Delta t}}&{\cdots}&{\hat{u}_{N,-N}^{+}\hat{\omega}_{N,-N}^{+}e^{i\hat{\omega}_{N,-N}^{+}\Delta t}}&{\hat{u}_{-N,-N}^{-}\hat{\omega}_{-N,-N}^{-}e^{i\hat{\omega}_{-N,-N}^{-}\Delta t}}&{\cdots}&{\hat{u}_{N,-N}^{-}\hat{\omega}_{N,-N}^{-}e^{i\hat{\omega}_{N,-N}^{-}\Delta t}}\\
{\vdots}&{\vdots}&{\vdots}&{\vdots}&{\vdots}&{\vdots}\\
{\hat{u}_{-N,N}^{+}\hat{\omega}_{-N,N}^{+}e^{i\hat{\omega}_{-N,N}^{+}\Delta t}}&{\cdots}&{\hat{u}_{N,N}^{+}\hat{\omega}_{N,N}^{+}e^{i\hat{\omega}_{N,N}^{+}\Delta t}}&{\hat{u}_{-N,N}^{-}\hat{\omega}_{-N,N}^{-}e^{i\hat{\omega}_{-N,N}^{-}\Delta t}}&{\cdots}&{\hat{u}_{N,N}^{-}\hat{\omega}_{N,N}^{-}e^{i\hat{\omega}_{N,N}^{-}\Delta t}}
\end{bmatrix}.
\end{equation}
\end{footnotesize}

$\mathbf{M_4}$ is a matrix of size $(4N+2)\times(4N+2)$ and given by:
\begin{equation}\label{Eq.C4}\tag{C4}
\mathbf{M_{4}}=
\begin{bmatrix}
diag(e^{i\tilde{\omega}_{-N}^{+}\Delta t}, \cdots, e^{i\tilde{\omega}_{N}^{+}\Delta t})&diag(e^{i\tilde{\omega}_{-N}^{-}\Delta t}, \cdots, e^{i\tilde{\omega}_{N}^{-}\Delta t})\\
diag(\tilde{\omega}_{-N}^{+}e^{i\tilde{\omega}_{-N}^{+}\Delta t}, \cdots, \tilde{\omega}_{N}^{+}e^{i\tilde{\omega}_{N}^{+}\Delta t})&diag(\tilde{\omega}_{-N}^{-}e^{i\tilde{\omega}_{-N}^{-}\Delta t}, \cdots, \tilde{\omega}_{N}^{-}e^{i\tilde{\omega}_{N}^{-}\Delta t})
\end{bmatrix}.
\end{equation}

\section{Details of FDTD simulations for temporal scattering behavior}
\label{App.D}
\renewcommand{\thefigure}{D.\arabic{figure}}
\setcounter{figure}{0}
The FDTD numerical simulations were performed by discretizing the space and time domains. In the simulations, we considered a space domain of length $X=100\pi$ m, i.e., $500\lambda_m$, and a time domain of length $T=80\pi$ s, i.e., $80T_m$/$800T_m$ for subsonic/supersonic modulation. The constitutive properties and modulation parameters were defined to be consistent with the settings in the analytical calculation. The subsonic and supersonic spatiotemporal modulation was both activated at $t_0=0.6T$, and deactivated at $t_1=0.65T$. By injecting wave packets with different central frequencies, we obtained the transient wavefields at different incident frequencies. The scattering spectra were then computed through fast Fourier transform (FFT). In addition, we obtained the wavefield evolution in the frequency-wavenumber domain through both spatial and temporal FFT (2D-FFT).

\printcredits

\bibliographystyle{cas-model2-names}

\bibliography{cas-refs}

\begin{thebibliography}{46}
\expandafter\ifx\csname natexlab\endcsname\relax\def\natexlab#1{#1}\fi
\providecommand{\url}[1]{\texttt{#1}}
\providecommand{\href}[2]{#2}
\providecommand{\path}[1]{#1}
\providecommand{\DOIprefix}{doi:}
\providecommand{\ArXivprefix}{arXiv:}
\providecommand{\URLprefix}{URL: }
\providecommand{\Pubmedprefix}{pmid:}
\providecommand{\doi}[1]{\href{http://dx.doi.org/#1}{\path{#1}}}
\providecommand{\Pubmed}[1]{\href{pmid:#1}{\path{#1}}}
\providecommand{\bibinfo}[2]{#2}
\ifx\xfnm\relax \def\xfnm[#1]{\unskip,\space#1}\fi
\bibitem[{Carminati et~al.(2021)Carminati, Chen, Pierrat and Shapiro}]{carminati2021universal}
\bibinfo{author}{Carminati, R.}, \bibinfo{author}{Chen, H.}, \bibinfo{author}{Pierrat, R.}, \bibinfo{author}{Shapiro, B.}, \bibinfo{year}{2021}.
\newblock \bibinfo{title}{Universal statistics of waves in a random time-varying medium}.
\newblock \bibinfo{journal}{Physical Review Letters} \bibinfo{volume}{127}, \bibinfo{pages}{094101}.
\bibitem[{Cassedy(1967)}]{cassedy1967dispersion}
\bibinfo{author}{Cassedy, E.S.}, \bibinfo{year}{1967}.
\newblock \bibinfo{title}{Dispersion relations in time-space periodic media part {II-Unstable interactions}}.
\newblock \bibinfo{journal}{Proceedings of the IEEE} \bibinfo{volume}{55}, \bibinfo{pages}{1154\--1168}.
\bibitem[{Cassedy and Oliner(1963)}]{cassedy1963dispersion}
\bibinfo{author}{Cassedy, E.S.}, \bibinfo{author}{Oliner, A.A.}, \bibinfo{year}{1963}.
\newblock \bibinfo{title}{Dispersion relations in time-space periodic media: part {I-Stable interactions}}.
\newblock \bibinfo{journal}{Proceedings of the IEEE} \bibinfo{volume}{51}, \bibinfo{pages}{1342\--1359}.
\bibitem[{Celli and Palermo(2024)}]{CELLI2024}
\bibinfo{author}{Celli, P.}, \bibinfo{author}{Palermo, A.}, \bibinfo{year}{2024}.
\newblock \bibinfo{title}{Time-modulated inerters as building blocks for nonreciprocal mechanical devices}.
\newblock \bibinfo{journal}{Journal of Sound and Vibration} \bibinfo{volume}{572}, \bibinfo{pages}{118178}.
\bibitem[{Chamanara et~al.(2017)Chamanara, Taravati, Deck-L{\'e}ger and Caloz}]{chamanara2017optical}
\bibinfo{author}{Chamanara, N.}, \bibinfo{author}{Taravati, S.}, \bibinfo{author}{Deck-L{\'e}ger, Z.L.}, \bibinfo{author}{Caloz, C.}, \bibinfo{year}{2017}.
\newblock \bibinfo{title}{Optical isolation based on space-time engineered asymmetric photonic band gaps}.
\newblock \bibinfo{journal}{Physical Review B} \bibinfo{volume}{96}, \bibinfo{pages}{155409}.
\bibitem[{Chen et~al.(2024)Chen, Mallejac, Bi, Xia and Fleury}]{chen2024experimental}
\bibinfo{author}{Chen, T.}, \bibinfo{author}{Mallejac, M.}, \bibinfo{author}{Bi, C.}, \bibinfo{author}{Xia, B.}, \bibinfo{author}{Fleury, R.}, \bibinfo{year}{2024}.
\newblock \bibinfo{title}{Experimental demonstration of a space-time modulated airborne acoustic circulator}.
\newblock \bibinfo{journal}{arXiv preprint arXiv:2411.16057} .
\bibitem[{Chen et~al.(2019)Chen, Li, Nassar, Norris, Daraio and Huang}]{chen2019nonreciprocal}
\bibinfo{author}{Chen, Y.}, \bibinfo{author}{Li, X.}, \bibinfo{author}{Nassar, H.}, \bibinfo{author}{Norris, A.N.}, \bibinfo{author}{Daraio, C.}, \bibinfo{author}{Huang, G.}, \bibinfo{year}{2019}.
\newblock \bibinfo{title}{Nonreciprocal wave propagation in a continuum-based metamaterial with space-time modulated resonators}.
\newblock \bibinfo{journal}{Physical Review Applied} \bibinfo{volume}{11}, \bibinfo{pages}{064052}.
\bibitem[{Delory et~al.(2024)Delory, Prada, Lanoy, Eddi, Fink and Lemoult}]{delory2024elastic}
\bibinfo{author}{Delory, A.}, \bibinfo{author}{Prada, C.}, \bibinfo{author}{Lanoy, M.}, \bibinfo{author}{Eddi, A.}, \bibinfo{author}{Fink, M.}, \bibinfo{author}{Lemoult, F.}, \bibinfo{year}{2024}.
\newblock \bibinfo{title}{Elastic wave packets crossing a space-time interface}.
\newblock \bibinfo{journal}{Physical Review Letters} \bibinfo{volume}{133}, \bibinfo{pages}{267201}.
\bibitem[{Dikopoltsev et~al.(2022)Dikopoltsev, Sharabi, Lyubarov, Lumer, Tsesses, Lustig, Kaminer and Segev}]{dikopoltsev2022light}
\bibinfo{author}{Dikopoltsev, A.}, \bibinfo{author}{Sharabi, Y.}, \bibinfo{author}{Lyubarov, M.}, \bibinfo{author}{Lumer, Y.}, \bibinfo{author}{Tsesses, S.}, \bibinfo{author}{Lustig, E.}, \bibinfo{author}{Kaminer, I.}, \bibinfo{author}{Segev, M.}, \bibinfo{year}{2022}.
\newblock \bibinfo{title}{Light emission by free electrons in photonic time-crystals}.
\newblock \bibinfo{journal}{Proceedings of the National Academy of Sciences} \bibinfo{volume}{119}, \bibinfo{pages}{e2119705119}.
\bibitem[{Engheta(2023)}]{engheta2023four}
\bibinfo{author}{Engheta, N.}, \bibinfo{year}{2023}.
\newblock \bibinfo{title}{Four-dimensional optics using time-varying metamaterials}.
\newblock \bibinfo{journal}{Science} \bibinfo{volume}{379}, \bibinfo{pages}{1190--1191}.
\bibitem[{Fleury et~al.(2015)Fleury, Sounas and Al{\`u}}]{fleury2015subwavelength}
\bibinfo{author}{Fleury, R.}, \bibinfo{author}{Sounas, D.L.}, \bibinfo{author}{Al{\`u}, A.}, \bibinfo{year}{2015}.
\newblock \bibinfo{title}{Subwavelength ultrasonic circulator based on spatiotemporal modulation}.
\newblock \bibinfo{journal}{Physical Review B} \bibinfo{volume}{91}, \bibinfo{pages}{174306}.
\bibitem[{Galiffi et~al.(2022)Galiffi, Tirole, Yin, Li, Vezzoli, Huidobro, Silveirinha, Sapienza, Al{\`u} and Pendry}]{galiffi2022photonics}
\bibinfo{author}{Galiffi, E.}, \bibinfo{author}{Tirole, R.}, \bibinfo{author}{Yin, S.}, \bibinfo{author}{Li, H.}, \bibinfo{author}{Vezzoli, S.}, \bibinfo{author}{Huidobro, P.A.}, \bibinfo{author}{Silveirinha, M.G.}, \bibinfo{author}{Sapienza, R.}, \bibinfo{author}{Al{\`u}, A.}, \bibinfo{author}{Pendry, J.B.}, \bibinfo{year}{2022}.
\newblock \bibinfo{title}{Photonics of time-varying media}.
\newblock \bibinfo{journal}{Advanced Photonics} \bibinfo{volume}{4}, \bibinfo{pages}{014002--014002}.
\bibitem[{Goldsberry et~al.(2020)Goldsberry, Wallen and Haberman}]{goldsberry2020nonreciprocal}
\bibinfo{author}{Goldsberry, B.M.}, \bibinfo{author}{Wallen, S.P.}, \bibinfo{author}{Haberman, M.R.}, \bibinfo{year}{2020}.
\newblock \bibinfo{title}{Nonreciprocal vibrations of finite elastic structures with spatiotemporally modulated material properties}.
\newblock \bibinfo{journal}{Physical Review B} \bibinfo{volume}{102}, \bibinfo{pages}{014312}.
\bibitem[{Horsley and Pendry(2023)}]{horsley2023quantum}
\bibinfo{author}{Horsley, S.A.}, \bibinfo{author}{Pendry, J.B.}, \bibinfo{year}{2023}.
\newblock \bibinfo{title}{Quantum electrodynamics of time-varying gratings}.
\newblock \bibinfo{journal}{Proceedings of the National Academy of Sciences} \bibinfo{volume}{120}, \bibinfo{pages}{e2302652120}.
\bibitem[{Jones et~al.(2024)Jones, Kildishev, Segev and Peroulis}]{jones2024time}
\bibinfo{author}{Jones, T.R.}, \bibinfo{author}{Kildishev, A.V.}, \bibinfo{author}{Segev, M.}, \bibinfo{author}{Peroulis, D.}, \bibinfo{year}{2024}.
\newblock \bibinfo{title}{Time-reflection of microwaves by a fast optically-controlled time-boundary}.
\newblock \bibinfo{journal}{Nature Communications} \bibinfo{volume}{15}, \bibinfo{pages}{6786}.
\bibitem[{Kim et~al.(2024)Kim, Chong and Daraio}]{kim2024temporal}
\bibinfo{author}{Kim, B.L.}, \bibinfo{author}{Chong, C.}, \bibinfo{author}{Daraio, C.}, \bibinfo{year}{2024}.
\newblock \bibinfo{title}{Temporal refraction in an acoustic phononic lattice}.
\newblock \bibinfo{journal}{Physical Review Letters} \bibinfo{volume}{133}, \bibinfo{pages}{077201}.
\bibitem[{Koutserimpas and Fleury(2018)}]{koutserimpas2018nonreciprocal}
\bibinfo{author}{Koutserimpas, T.T.}, \bibinfo{author}{Fleury, R.}, \bibinfo{year}{2018}.
\newblock \bibinfo{title}{Nonreciprocal gain in non-hermitian time-floquet systems}.
\newblock \bibinfo{journal}{Physical Review Letters} \bibinfo{volume}{120}, \bibinfo{pages}{087401}.
\bibitem[{Lee et~al.(2018)Lee, Son, Park, Kang, Jeon, Rotermund and Min}]{lee2018linear}
\bibinfo{author}{Lee, K.}, \bibinfo{author}{Son, J.}, \bibinfo{author}{Park, J.}, \bibinfo{author}{Kang, B.}, \bibinfo{author}{Jeon, W.}, \bibinfo{author}{Rotermund, F.}, \bibinfo{author}{Min, B.}, \bibinfo{year}{2018}.
\newblock \bibinfo{title}{Linear frequency conversion via sudden merging of meta-atoms in time-variant metasurfaces}.
\newblock \bibinfo{journal}{Nature Photonics} \bibinfo{volume}{12}, \bibinfo{pages}{765--773}.
\bibitem[{Li et~al.(2019a)Li, Shen, Zhu, Xie and Cummer}]{li2019nonreciprocal}
\bibinfo{author}{Li, J.}, \bibinfo{author}{Shen, C.}, \bibinfo{author}{Zhu, X.}, \bibinfo{author}{Xie, Y.}, \bibinfo{author}{Cummer, S.A.}, \bibinfo{year}{2019}a.
\newblock \bibinfo{title}{Nonreciprocal sound propagation in space-time modulated media}.
\newblock \bibinfo{journal}{Physical Review B} \bibinfo{volume}{99}, \bibinfo{pages}{144311}.
\bibitem[{Li et~al.(2019b)Li, Ni, Weiner, Al{\`u} and Khanikaev}]{li2019topological}
\bibinfo{author}{Li, M.}, \bibinfo{author}{Ni, X.}, \bibinfo{author}{Weiner, M.}, \bibinfo{author}{Al{\`u}, A.}, \bibinfo{author}{Khanikaev, A.B.}, \bibinfo{year}{2019}b.
\newblock \bibinfo{title}{Topological phases and nonreciprocal edge states in non-hermitian floquet insulators}.
\newblock \bibinfo{journal}{Physical Review B} \bibinfo{volume}{100}, \bibinfo{pages}{045423}.
\bibitem[{Lyubarov et~al.(2022)Lyubarov, Lumer, Dikopoltsev, Lustig, Sharabi and Segev}]{lyubarov2022amplified}
\bibinfo{author}{Lyubarov, M.}, \bibinfo{author}{Lumer, Y.}, \bibinfo{author}{Dikopoltsev, A.}, \bibinfo{author}{Lustig, E.}, \bibinfo{author}{Sharabi, Y.}, \bibinfo{author}{Segev, M.}, \bibinfo{year}{2022}.
\newblock \bibinfo{title}{Amplified emission and lasing in photonic time crystals}.
\newblock \bibinfo{journal}{Science} \bibinfo{volume}{377}, \bibinfo{pages}{425--428}.
\bibitem[{Marconi et~al.(2020)Marconi, Riva, Di~Ronco, Cazzulani, Braghin and Ruzzene}]{Marconi2020experimental}
\bibinfo{author}{Marconi, J.}, \bibinfo{author}{Riva, E.}, \bibinfo{author}{Di~Ronco, M.}, \bibinfo{author}{Cazzulani, G.}, \bibinfo{author}{Braghin, F.}, \bibinfo{author}{Ruzzene, M.}, \bibinfo{year}{2020}.
\newblock \bibinfo{title}{Experimental observation of nonreciprocal band gaps in a space-time-modulated beam using a shunted piezoelectric array}.
\newblock \bibinfo{journal}{Physical Review Applied} \bibinfo{volume}{13}, \bibinfo{pages}{031001}.
\bibitem[{Miyamaru et~al.(2021)Miyamaru, Mizuo, Nakanishi, Nakata, Hasebe, Nagase, Matsubara, Goto, P{\'e}rez-Urquizo, Mad{\'e}o et~al.}]{miyamaru2021ultrafast}
\bibinfo{author}{Miyamaru, F.}, \bibinfo{author}{Mizuo, C.}, \bibinfo{author}{Nakanishi, T.}, \bibinfo{author}{Nakata, Y.}, \bibinfo{author}{Hasebe, K.}, \bibinfo{author}{Nagase, S.}, \bibinfo{author}{Matsubara, Y.}, \bibinfo{author}{Goto, Y.}, \bibinfo{author}{P{\'e}rez-Urquizo, J.}, \bibinfo{author}{Mad{\'e}o, J.}, et~al., \bibinfo{year}{2021}.
\newblock \bibinfo{title}{Ultrafast frequency-shift dynamics at temporal boundary induced by structural-dispersion switching of waveguides}.
\newblock \bibinfo{journal}{Physical Review Letters} \bibinfo{volume}{127}, \bibinfo{pages}{053902}.
\bibitem[{Moussa et~al.(2023)Moussa, Xu, Yin, Galiffi, Ra’di and Al{\`u}}]{Moussa2023observation}
\bibinfo{author}{Moussa, H.}, \bibinfo{author}{Xu, G.}, \bibinfo{author}{Yin, S.}, \bibinfo{author}{Galiffi, E.}, \bibinfo{author}{Ra’di, Y.}, \bibinfo{author}{Al{\`u}, A.}, \bibinfo{year}{2023}.
\newblock \bibinfo{title}{Observation of temporal reflection and broadband frequency translation at photonic time interfaces}.
\newblock \bibinfo{journal}{Nature Physics} \bibinfo{volume}{19}, \bibinfo{pages}{863\--868}.
\bibitem[{Nassar et~al.(2017a)Nassar, Chen, Norris and Huang}]{nassar2017non}
\bibinfo{author}{Nassar, H.}, \bibinfo{author}{Chen, H.}, \bibinfo{author}{Norris, A.}, \bibinfo{author}{Huang, G.}, \bibinfo{year}{2017}a.
\newblock \bibinfo{title}{Non-reciprocal flexural wave propagation in a modulated metabeam}.
\newblock \bibinfo{journal}{Extreme Mechanics Letters} \bibinfo{volume}{15}, \bibinfo{pages}{97--102}.
\bibitem[{Nassar et~al.(2017b)Nassar, Xu, Norris and Huang}]{nassar2017modulated}
\bibinfo{author}{Nassar, H.}, \bibinfo{author}{Xu, X.}, \bibinfo{author}{Norris, A.}, \bibinfo{author}{Huang, G.}, \bibinfo{year}{2017}b.
\newblock \bibinfo{title}{Modulated phononic crystals: Non-reciprocal wave propagation and willis materials}.
\newblock \bibinfo{journal}{Journal of the Mechanics and Physics of Solids} \bibinfo{volume}{101}, \bibinfo{pages}{10\--29}.
\bibitem[{Nassar et~al.(2020)Nassar, Yousefzadeh, Fleury, Ruzzene, Al{\`u}, Daraio, Norris, Huang and Haberman}]{nassar2020nonreciprocity}
\bibinfo{author}{Nassar, H.}, \bibinfo{author}{Yousefzadeh, B.}, \bibinfo{author}{Fleury, R.}, \bibinfo{author}{Ruzzene, M.}, \bibinfo{author}{Al{\`u}, A.}, \bibinfo{author}{Daraio, C.}, \bibinfo{author}{Norris, A.N.}, \bibinfo{author}{Huang, G.}, \bibinfo{author}{Haberman, M.R.}, \bibinfo{year}{2020}.
\newblock \bibinfo{title}{Nonreciprocity in acoustic and elastic materials}.
\newblock \bibinfo{journal}{Nature Reviews Materials} \bibinfo{volume}{5}, \bibinfo{pages}{667--685}.
\bibitem[{Ni et~al.(2021)Ni, Kim and Al{\`u}}]{ni2021topological}
\bibinfo{author}{Ni, X.}, \bibinfo{author}{Kim, S.}, \bibinfo{author}{Al{\`u}, A.}, \bibinfo{year}{2021}.
\newblock \bibinfo{title}{Topological insulator in two synthetic dimensions based on an optomechanical resonator}.
\newblock \bibinfo{journal}{Optica} \bibinfo{volume}{8}, \bibinfo{pages}{1024--1032}.
\bibitem[{Palermo et~al.(2020)Palermo, Celli, Yousefzadeh, Daraio and Marzani}]{palermo2020surface}
\bibinfo{author}{Palermo, A.}, \bibinfo{author}{Celli, P.}, \bibinfo{author}{Yousefzadeh, B.}, \bibinfo{author}{Daraio, C.}, \bibinfo{author}{Marzani, A.}, \bibinfo{year}{2020}.
\newblock \bibinfo{title}{Surface wave non-reciprocity via time-modulated metamaterials}.
\newblock \bibinfo{journal}{Journal of the Mechanics and Physics of Solids} \bibinfo{volume}{145}, \bibinfo{pages}{104181}.
\bibitem[{Pendry et~al.(2021)Pendry, Galiffi and Huidobro}]{pendry2021gain}
\bibinfo{author}{Pendry, J.}, \bibinfo{author}{Galiffi, E.}, \bibinfo{author}{Huidobro, P.}, \bibinfo{year}{2021}.
\newblock \bibinfo{title}{Gain mechanism in time-dependent media}.
\newblock \bibinfo{journal}{Optica} \bibinfo{volume}{8}, \bibinfo{pages}{636--637}.
\bibitem[{Ramaccia et~al.(2019)Ramaccia, Sounas, Alu, Toscano and Bilotti}]{ramaccia2019phase}
\bibinfo{author}{Ramaccia, D.}, \bibinfo{author}{Sounas, D.L.}, \bibinfo{author}{Alu, A.}, \bibinfo{author}{Toscano, A.}, \bibinfo{author}{Bilotti, F.}, \bibinfo{year}{2019}.
\newblock \bibinfo{title}{Phase-induced frequency conversion and doppler effect with time-modulated metasurfaces}.
\newblock \bibinfo{journal}{IEEE Transactions on Antennas and Propagation} \bibinfo{volume}{68}, \bibinfo{pages}{1607--1617}.
\bibitem[{Santini and Riva(2023)}]{santini2023elastic}
\bibinfo{author}{Santini, J.}, \bibinfo{author}{Riva, E.}, \bibinfo{year}{2023}.
\newblock \bibinfo{title}{Elastic temporal waveguiding}.
\newblock \bibinfo{journal}{New Journal of Physics} \bibinfo{volume}{25}, \bibinfo{pages}{013031}.
\bibitem[{Sounas and Al{\`u}(2017)}]{sounas2017non}
\bibinfo{author}{Sounas, D.L.}, \bibinfo{author}{Al{\`u}, A.}, \bibinfo{year}{2017}.
\newblock \bibinfo{title}{Non-reciprocal photonics based on time modulation}.
\newblock \bibinfo{journal}{Nature Photonics} \bibinfo{volume}{11}, \bibinfo{pages}{774--783}.
\bibitem[{Trainiti and Ruzzene(2016)}]{trainiti2016non}
\bibinfo{author}{Trainiti, G.}, \bibinfo{author}{Ruzzene, M.}, \bibinfo{year}{2016}.
\newblock \bibinfo{title}{Non-reciprocal elastic wave propagation in spatiotemporal periodic structures}.
\newblock \bibinfo{journal}{New Journal of Physics} \bibinfo{volume}{18}, \bibinfo{pages}{083047}.
\bibitem[{Trainiti et~al.(2019)Trainiti, Xia, Marconi, Cazzulani, Erturk and Ruzzene}]{Trainiti2019time}
\bibinfo{author}{Trainiti, G.}, \bibinfo{author}{Xia, Y.}, \bibinfo{author}{Marconi, J.}, \bibinfo{author}{Cazzulani, G.}, \bibinfo{author}{Erturk, A.}, \bibinfo{author}{Ruzzene, M.}, \bibinfo{year}{2019}.
\newblock \bibinfo{title}{Time-periodic stiffness modulation in elastic metamaterials for selective wave filtering: Theory and experiment}.
\newblock \bibinfo{journal}{Physical Review Letters} \bibinfo{volume}{122}, \bibinfo{pages}{124301}.
\bibitem[{V{\'a}zquez-Lozano and Liberal(2023)}]{vazquez2023incandescent}
\bibinfo{author}{V{\'a}zquez-Lozano, J.E.}, \bibinfo{author}{Liberal, I.}, \bibinfo{year}{2023}.
\newblock \bibinfo{title}{Incandescent temporal metamaterials}.
\newblock \bibinfo{journal}{Nature Communications} \bibinfo{volume}{14}, \bibinfo{pages}{4606}.
\bibitem[{Wang et~al.(2021)Wang, Dutt, Wojcik and Fan}]{wang2021topological}
\bibinfo{author}{Wang, K.}, \bibinfo{author}{Dutt, A.}, \bibinfo{author}{Wojcik, C.C.}, \bibinfo{author}{Fan, S.}, \bibinfo{year}{2021}.
\newblock \bibinfo{title}{Topological complex-energy braiding of non-hermitian bands}.
\newblock \bibinfo{journal}{Nature} \bibinfo{volume}{598}, \bibinfo{pages}{59--64}.
\bibitem[{Wang et~al.(2025)Wang, Shao, Chen, Chen, Qian, Wu, Duan, Alu and Huang}]{wang2025temporal}
\bibinfo{author}{Wang, S.}, \bibinfo{author}{Shao, N.}, \bibinfo{author}{Chen, H.}, \bibinfo{author}{Chen, J.}, \bibinfo{author}{Qian, H.}, \bibinfo{author}{Wu, Q.}, \bibinfo{author}{Duan, H.}, \bibinfo{author}{Alu, A.}, \bibinfo{author}{Huang, G.}, \bibinfo{year}{2025}.
\newblock \bibinfo{title}{Temporal refraction and reflection in modulated mechanical metabeams: theory and physical observation}.
\newblock \bibinfo{journal}{arXiv preprint arXiv:2501.09989} .
\bibitem[{Wang et~al.(2018)Wang, Yousefzadeh, Chen, Nassar, Huang and Daraio}]{wang2018observation}
\bibinfo{author}{Wang, Y.}, \bibinfo{author}{Yousefzadeh, B.}, \bibinfo{author}{Chen, H.}, \bibinfo{author}{Nassar, H.}, \bibinfo{author}{Huang, G.}, \bibinfo{author}{Daraio, C.}, \bibinfo{year}{2018}.
\newblock \bibinfo{title}{Observation of nonreciprocal wave propagation in a dynamic phononic lattice}.
\newblock \bibinfo{journal}{Physical Review Letters} \bibinfo{volume}{121}, \bibinfo{pages}{194301}.
\bibitem[{Wapenaar(2025)}]{wapenaar2025green}
\bibinfo{author}{Wapenaar, K.}, \bibinfo{year}{2025}.
\newblock \bibinfo{title}{Green’s functions, propagation invariants, reciprocity theorems, wave-field representations and propagator matrices in two-dimensional time-dependent materials}, in: \bibinfo{booktitle}{Proceedings A}, \bibinfo{organization}{The Royal Society}. p. \bibinfo{pages}{20240479}.
\bibitem[{Wapenaar et~al.(2024)Wapenaar, Aichele and van Manen}]{wapenaar2024waves}
\bibinfo{author}{Wapenaar, K.}, \bibinfo{author}{Aichele, J.}, \bibinfo{author}{van Manen, D.J.}, \bibinfo{year}{2024}.
\newblock \bibinfo{title}{Waves in space-dependent and time-dependent materials: A systematic comparison}.
\newblock \bibinfo{journal}{Wave Motion} \bibinfo{volume}{130}, \bibinfo{pages}{103374}.
\bibitem[{Wu et~al.(2021)Wu, Chen, Nassar and Huang}]{wu2021non}
\bibinfo{author}{Wu, Q.}, \bibinfo{author}{Chen, H.}, \bibinfo{author}{Nassar, H.}, \bibinfo{author}{Huang, G.}, \bibinfo{year}{2021}.
\newblock \bibinfo{title}{Non-reciprocal rayleigh wave propagation in space--time modulated surface}.
\newblock \bibinfo{journal}{Journal of the Mechanics and Physics of Solids} \bibinfo{volume}{146}, \bibinfo{pages}{104196}.
\bibitem[{Wu and Grbic(2019)}]{wu2019serrodyne}
\bibinfo{author}{Wu, Z.}, \bibinfo{author}{Grbic, A.}, \bibinfo{year}{2019}.
\newblock \bibinfo{title}{Serrodyne frequency translation using time-modulated metasurfaces}.
\newblock \bibinfo{journal}{IEEE Transactions on Antennas and Propagation} \bibinfo{volume}{68}, \bibinfo{pages}{1599--1606}.
\bibitem[{Xia et~al.(2021)Xia, Riva, Rosa, Cazzulani, Erturk, Braghin and Ruzzene}]{Xia2021experimental}
\bibinfo{author}{Xia, Y.}, \bibinfo{author}{Riva, E.}, \bibinfo{author}{Rosa, M.I.}, \bibinfo{author}{Cazzulani, G.}, \bibinfo{author}{Erturk, A.}, \bibinfo{author}{Braghin, F.}, \bibinfo{author}{Ruzzene, M.}, \bibinfo{year}{2021}.
\newblock \bibinfo{title}{Experimental observation of temporal pumping in electromechanical waveguides}.
\newblock \bibinfo{journal}{Physical Review Letters} \bibinfo{volume}{126}, \bibinfo{pages}{095501}.
\bibitem[{Xiao et~al.(2014)Xiao, Maywar and Agrawal}]{xiao2014reflection}
\bibinfo{author}{Xiao, Y.}, \bibinfo{author}{Maywar, D.N.}, \bibinfo{author}{Agrawal, G.P.}, \bibinfo{year}{2014}.
\newblock \bibinfo{title}{Reflection and transmission of electromagnetic waves at a temporal boundary}.
\newblock \bibinfo{journal}{Optics letters} \bibinfo{volume}{39}, \bibinfo{pages}{574--577}.
\bibitem[{Zhou et~al.(2020)Zhou, Alam, Karimi, Upham, Reshef, Liu, Willner and Boyd}]{zhou2020broadband}
\bibinfo{author}{Zhou, Y.}, \bibinfo{author}{Alam, M.Z.}, \bibinfo{author}{Karimi, M.}, \bibinfo{author}{Upham, J.}, \bibinfo{author}{Reshef, O.}, \bibinfo{author}{Liu, C.}, \bibinfo{author}{Willner, A.E.}, \bibinfo{author}{Boyd, R.W.}, \bibinfo{year}{2020}.
\newblock \bibinfo{title}{Broadband frequency translation through time refraction in an epsilon-near-zero material}.
\newblock \bibinfo{journal}{Nature communications} \bibinfo{volume}{11}, \bibinfo{pages}{2180}.

\end{thebibliography}


\end{document}